\documentclass[12pt]{article}
\usepackage[a4paper, hmargin=2cm, vmargin=2.5cm]{geometry}
\usepackage{amsthm,amsmath,amsfonts, amssymb, multirow, lscape}
\usepackage[authoryear]{natbib} \bibpunct{(}{)}{,}{a}{,}{,}
\usepackage{graphicx}        

\newcommand{\btheta}  {\boldsymbol\theta}
\newcommand{\bmu}     {\boldsymbol\mu}
\newcommand{\btau}    {\boldsymbol\tau}
\newcommand{\bdelta}  {\boldsymbol\delta}
\newcommand{\bPsi}    {\boldsymbol\Psi}
\newcommand{\bLambda} {\boldsymbol\Lambda}
\newcommand{\bGamma}  {\boldsymbol\Gamma}
\newcommand{\bSigma}  {\boldsymbol\Sigma}
\newcommand{\bDelta}  {\boldsymbol\Delta}

\newcommand{\bu}	  {\mbox{\boldmath $u$}}
\newcommand{\by}	  {\mbox{\boldmath $y$}}
\newcommand{\bY}	  {\mbox{\boldmath $Y$}}
\newcommand{\bD}	  {\mbox{\boldmath $D$}}
\newcommand{\bI}	  {\mbox{\boldmath $I$}}
\newcommand{\bR}	  {\mbox{\boldmath $R$}}
\newcommand{\bzero}	  {\mbox{\boldmath $0$}}

\begin{document}

\title{On Mixtures of Skew Normal and Skew $t$-Distributions}

\author{Sharon X. Lee, Geoffrey J. McLachlan	 \\ 
Department of mathematics, the University of Queensland,\\ Brisbane, Australia}
\date{}

\maketitle


\begin{abstract}
Finite mixtures of skew distributions have emerged as an effective tool
in modelling heterogeneous data with asymmetric features.
With various proposals appearing rapidly in the recent years,
which are similar but not identical, 
the connections between them and their relative performance
becomes rather unclear.
This paper aims to provide a concise overview of
these developments by presenting a systematic classification
of the existing skew symmetric distributions into four types,
thereby clarifying their close relationships.
This also aids in understanding the link between
some of the proposed expectation-maximization (EM)
based algorithms for the computation of the maximum likelihood 
estimates of the parameters of the models.
The final part of this paper presents an illustration 
of the performance of these mixture models 
in clustering a real dataset, relative to other
non-elliptically contoured clustering methods
and associated algorithms for their implementation.
\end{abstract}

\section{Introduction}
\label{intro}

In recent years, non-normal distributions have received
substantial interest in the statistics literature.  
The growing need for more flexible tools to analyze datasets
that exhibit non-normal features, including asymmetry,
multimodality, and heavy tails, has led to intense
development in non-normal model-based methods.  
In particular, finite mixtures of skew distributions
have emerged as a promising alternative to the traditional
Gaussian mixture modelling. They have been successfully
applied to numerous datasets from a wide range of fields,
including the medical sciences, bioinformatics, environmetrics,
engineering, economics, and financial sciences.  
Some recent applications of multivariate skew normal
and skew $t$-mixture models include \citet{J004, J108, J119}, 
and \citet{J118}.  

The rich literature and active discussion of skew distributions
was initiated by the pioneering work of \citet{J005},
in which the univariate skew normal distribution was introduced.
Following its generalization to the multivariate
skew normal distribution in \citet{J001},
the number of contributions have grown rapidly.
The concept of introducing additional parameters to
regulate skewness in a distribution was subsequently
extended to other parametric families,
yielding the skew elliptical family;
for a comprehensive survey of skew distributions, see, for example,
the articles by \citet{J021, J017, J071}, and also the book
edited by \citet{B001}.

Besides the skew normal distribution, which plays a central role
in these developments, the skew $t$-distribution has also received
much attention. Being a natural extension of the $t$-distribution,
the skew $t$-distribution retains reasonable tractability and
is more robust against outliers than the skew normal distribution.
Finite mixtures of skew normal and skew $t$-distributions
have been studied by several authors, including
\citet{J026, J046, J004, J047, J028, J027, J066, J077, J103},
and \citet{J117}, among others.
With the existence of so many proposals,
with their various characterizations of skew normal
 and skew $t$-distributions, it becomes rather unclear
 how these proposals are related to each other, and
to what extent can the subtle differences between them
have in practical applications.
    
This paper provides a concise overview of various recent developments
of mixtures of skew normal and skew $t$-distributions.
An illustration is given of the performance of mixtures of these distributions
and some other skew mixture models in clustering a real dataset.
We first present a systematic classification of
multivariate skew normal and skew $t$-distributions,
with special references to those used in various
existing proposals of finite mixture models.
We then illustrate the relative performance of these models
and other related algorithms by applications to a real dataset.   

Recently, \citet{J068} referred to the 
skew normal and skew $t$-distributions
of \citet{J004} as `restricted' skew distributions,
and the class of skew elliptical distributions of \citet{J002}
as having the `unrestricted' form.
While this terminology was later briefly discussed in \citet{J103}
when outlining the equivalence between the skew distributions of
\citet{J001}, \citet{J004}, and \citet{J047},
further details were not given. This papers aims to fill this gap.
We shall adopt the above terminology,
and expand this idea further to classify more general forms of
skew distributions, namely, the `extended' and `generalized' forms.

The remainder of this paper is organized as follows.
In Section \ref{sec:classif}, we present the classification scheme
for multivariate skew normal and skew $t$-distributions,
clarifying the connections between various variants.
Next, we discuss the development of currently available algorithms
for fitting mixtures of multivariate skew normal and skew $t$-distributions
in Section \ref{sec:FM}, pointing out the equivalence between
some of these algorithms.
Section \ref{sec:DLBCL}
presents an application to automated flow cytometric analysis,
and comparisons are made with the results of other
model-based clustering methods.
Finally, some concluding remarks are given in Section \ref{sec:concl}.

\section{Classification of multivariate skew normal and skew $t$-distributions}
\label{sec:classif}

\subsection{Multivariate skew normal distributions}
\label{sec:MSN}

Since the seminal article by \citet{J001} on the
multivariate skew normal (MSN) distribution,
numerous `extensions' of the so-called skew normal distribution
have appeared in rapid succession. The number of contributions
are now so many that it is beyond the scope of this paper
to include them all here.
However, most of these developments can be considered as
special cases of the fundamental skew normal (FUSN) distribution
\citep{J008}, and can be systematically classified into four types,
namely, the restricted, unrestricted, extended, and generalized forms.

We begin by briefly discussing the FUSN distribution,
since it encompasses the first three forms of MSN distributions.
The FUSN distribution itself is a generalized form of the MSN distribution.
It can be generated by conditioning a multivariate normal variable
on another (univariate or multivariate) random variable.
Suppose $\bY_1 \sim N_p(\bzero, \bSigma)$ and $\bY_0$ is a
$q$-dimensional random vector.
Adopting the notation as used in \citet{J001}, we let \linebreak
 $\bY_1 \mid \bY_0 > \bzero$ be the vector $\bY_1$
if all elements of $\bY_0$ are positive and $- \bY_1$ otherwise.
Then $\bY = \bmu + (\bY_1 \mid \bY_0 + \btau > \bzero)$
has a FUSN distribution.
It is important to note that $\bY_0$ is \emph{not} necessarily
normally distributed, but in the restricted, unrestricted,
and extended cases, it is restricted to be a random normal variate.
The parameter $\btau \in \mathbb{R}^q$, known as the extension parameter,
can be viewed as a location shift for the latent variable $\bY_0$.
When the joint distribution of $\bY_1$ and $\bY_0$ is multivariate normal,
the FUSN distribution reduces to a location-scale variant of the 
canonical FUSN (CFUSN) distribution, given by
\begin{equation}
\bY = (\bY_1 \mid \bY_0 > \bzero),
\label{eq2.16}
\end{equation}
where
\begin{equation}
\left[\begin{array}{c}\bY_0\\\bY_1\end{array}\right]
        \sim N_{q+p} \left(\left[\begin{array}{c}\btau\\
        \bmu\end{array}\right],
        \left[\begin{array}{cc}\bGamma & \bDelta^T \\
        \bDelta    &    \bSigma\end{array}\right]\right),
        \label{eq2.17}
\end{equation}
where $\btau$ is a $q$-dimensional vector, $\bmu$ is $p$-dimensional vector,
$\bGamma$ is a $q\times q$ scale matrix,
$\bDelta$ is an arbitrary $p \times q$ matrix,
and $\bSigma$ is a $q \times q$ scale matrix.

The restricted case corresponds to a highly specialized form
of (\ref{eq2.17}), where $\bY_0$ is restricted to be univariate
(that is, $q=1$), $\btau=0$, and $\bGamma = 1$.
In the unrestricted case, both $\bY_0$ and $\bY_1$
have a $p$-dimensional normal distribution (that is, $q$=$p$).
Note that the use of ``restricted'' here 
refers to restrictions on the random vector in the (conditioning-type)
stochastic definition of the skew distribution.  It is not a restriction
on the parameter space, and so a ``restricted'' form of a skew
distribution is not necessarily nested within its corrresponding
``unrestricted'' form.  Indeed, the restricted
and unrestricted forms coincide in the univariate case.

The extended form has no restriction on the dimensions of $\bY_0$,
but $\btau$ can be a non-zero vector.
When $\bY_0$ is not normally distributed,
the density of $\bY$ has the generalized form.
A summary of some of the existing multivariate
skew normal distributions is given in Tables~\ref{tab:2.1} and~\ref{tab:2.2},
where rMSN , uMSN, eMSN, and gMSN refer to the restricted,
unrestricted, extended, and generalized version, respectively,
of the multivariate skew normal distribution.
The list is not exhaustive, and the names appearing in
the final columns are representative examples only.

\begin{table} \footnotesize
    \centering
        \begin{tabular}{|c|c|c|c|}
            \hline
            Case    &    Notation    &    Restrictions on FUSN    &    Examples \\
            \hline                    
            \hline
                restricted    &    rMSN    &    $q=1$, $\tau=0$,
                and $\left[\begin{array}{c}Y_0 \\ \bY_1\end{array}\right]
                \sim N_{1+p}$
                        &    A-MSN, B-MSN, SNI-SN, P-MSN \\
            \hline
                unrestricted    &    uMSN    & $q=p$, $\btau=\bzero$, and
                $\left[\begin{array}{c} \bY_0 \\ \bY_1\end{array}\right]
                \sim N_{2p}$
                        & S-MSN, G-MSN \\
            \hline
                extended    &    eMSN    &    $\btau \neq \bzero$, and
                $\left[\begin{array}{c} \bY_0 \\ \bY_1\end{array}\right]
                \sim N_{q+p}$
                        &    ESN, CSN, HSN, SUN \\
            \hline
                generalized    &    gMSN    & $\bY_1$ is normally distributed
                & FUSN, GSN, FSN, SMSN \\
            \hline
        \end{tabular}
    \caption{Classification of multivariate skew normal distributions.}
    \label{tab:2.1}
\end{table}

\begin{table}
    \centering
    \begin{tabular}{|c|c|c|}
        \hline
        Abbreviation    &    Name    &    References    \\        
        \hline
        \multicolumn{1}{|c}{\textbf{rMSN}}    & \multicolumn{1}{c}{}     &    \\
        \hline
        A-MSN    &    Azzalini's MSN    &    \citet{J001}    \\
        B-MSN    &    Branco's MSN    	&    \citet{J012}    \\
        SNI-SN   &    skew normal/independent    MSN&    \citet{J049} \\
        P-MSN    &    Pyne's MSN    	&    \citet{J004}    \\    
        \hline
        \multicolumn{1}{|c}{\textbf{uMSN}}    &    \multicolumn{1}{c}{} &    \\
        \hline
        S-MSN    &    Sahu's MSN    	&    \citet{J002}    \\
        G-MSN    &    Gupta's MSN    	&    \citet{J014}    \\
        \hline
        \multicolumn{1}{|c}{\textbf{eMSN}}    & \multicolumn{1}{c}{}     &    \\
        \hline
        ESN    &    Extended MSN    	&    \citet{J100}   \\
        CSN    &    Closed    MSN    	&    \citet{J018}   \\
        HSN    &    Hierarchical MSN 	&    \citet{J015} 	\\
        SUN    &    Unified MSN    		&    \citet{J017} 	\\
        \hline        
        \multicolumn{1}{|c}{\textbf{gMSN}}    & \multicolumn{1}{c}{}     &    \\
        \hline
        FUSN    &    Fundamental MSN    	&    \citet{J008}   \\
        GSN     &    Generalized MSN    	&    \citet{J023}   \\
        FSN     &    Flexible MSN        	&    \citet{J072}   \\
        SMSN    &    Shape mixture of MSN  	&    \citet{J076} 	\\
        \hline                    
    \end{tabular}
    \caption{Summary of the abbreviations of skew normal distributions
    used in Table \ref{tab:2.1}.}
    \label{tab:2.2}
\end{table}

\subsubsection{Restricted multivariate skew normal distributions}
\label{sec:rMSN}

The restricted case is one of the simplest multivariate forms of
the FUSN distribution. The latent variable $Y_0$ is assumed to be
a univariate random normal variable, and its correlation with $\bY_1$
is controlled by $\bdelta \in \mathbb{R}^p$.
There exists two parallel forms of stochastic representation
for a MSN random variable, obtained via the conditioning and
convolution mechanism \citep{J021}.  
In general, the conditioning-type stochastic representation
of a restricted MSN (rMSN) distribution is given by
\begin{equation}
\bY = \bmu + (\bY_1 \mid Y_0 > 0),
\label{eq2.18}
\end{equation}
where
\begin{equation}
\left[\begin{array}{c} Y_0 \\ \bY_1 \end{array}\right]
        \sim N_{1+p} \left(\left[\begin{array}{c} 0 \\
        \bzero \end{array}\right],
        \left[\begin{array}{cc} 1 & {\bdelta}^T \\
        \bdelta & \bSigma
        \end{array}\right]\right).
\label{eq2.19}
\end{equation}

Alternatively, the rMSN distribution can be generated via
the convolution approach, which leads to a convolution-type
stochastic representation, given by
\begin{equation}
\bY = \bmu + \tilde{\bdelta}\left|\tilde{Y}_0\right|
        + \tilde{\bY}_1,
\label{eq2.20}
\end{equation}
where $\tilde{Y}_0 \sim N_1(0,1)$ and
$\tilde{\bY}_1 \sim N_p(\bzero, \tilde{\bSigma})$ are independent,
and where $|\tilde{\bY}_0|$ denotes the vector whose $i$th element 
is given by the absolute value of the $i$th element of $\tilde{\bY}_0$.
Note that the parameters in (\ref{eq2.20}) are not identical
to those in (\ref{eq2.18}) and (\ref{eq2.19}),
but can be obtained from the latter
by taking $\tilde{\bdelta} = \bdelta$
and $\tilde{\bSigma} = \bSigma - \bdelta\bdelta^T$.
The connection between the pairs $(\bdelta, \bSigma)$ and
$(\tilde{\bdelta}, \tilde{\bSigma})$,
are discussed in more detail in \citet{J100}.
The skew normal distribution proposed by \citet{J001, J012, J049},
and \citet{J004} are identical after reparameterization,
and can be formulated as the rMSN distribution.
\newline

\noindent
\textbf{The first multivariate skew normal distribution (A-MSN)} \newline
The first formal definition of the univariate skew normal distribution
dates back to \citet{J005}. However, its extension to
the multivariate case
did not appear until just over a decade later.
The widely accepted `original' multivariate skew normal distribution
was introduced by \citet{J001}. The density of this distribution,
denoted by A-MSN$(\bmu, \bSigma, \bdelta_A)$ (with some changes in notation)
takes the form
\begin{equation}
f(\by; \bmu, \bSigma, \bdelta_A) = 2 \phi_p(\by; \bmu, \bSigma)
        \Phi_1(\bdelta_A^T\bR^{-1}\bD^{-1}(\by -\bmu); 0, 1 - \bdelta_A^T\bR^{-1}\bdelta_A),
\label{AzzaSN}
\end{equation}
where 
$\bR = \bD^{-1} \bSigma \bD^{-1}$ is the correlation matrix, 
$\bD = \mbox{diag}(\sqrt{\Sigma}_{11}, \ldots, \sqrt{\Sigma}_{pp})$
is a diagonal matrix formed by extracting the diagonal elements
of $\bSigma$, and $\Sigma_{ij}$ denotes the $ij$th entry of $\bSigma$.
We let $\phi_p(.;\bmu, \bSigma)$ be the $p$-dimensional
normal density with mean $\bmu$ and covariance matrix $\bSigma$,
and $\Phi_1(.;\mu, \sigma^2)$ is the (univariate) normal
distribution function of a normal variable with mean $\mu$
and variance $\sigma^2$.
To avoid ambiguity in the notation, we have appended a subscript
to some of the parameters used in different versions 
of the rMSN distributions throughout this paper,
for example, $\bdelta_A$ denotes the version of $\bdelta$ 
used in the A-MSN distribution. 
The density (\ref{AzzaSN}) was obtained via the conditioning method (\ref{eq2.18}),
with $\bY = \bmu + \bD(\bY_1 \mid Y_0 > 0)$,
where $Y_0$ and $\bY_1$ are distributed according to (\ref{eq2.19}).
It corresponds to the rMSN distribution in (\ref{eq2.19})
with $\bdelta$ replaced by $\bD\bdelta_A$.
This characterization of the MSN distribution was adopted
in the work of \citet{J028} when formulating finite mixtures
of skew normal distributions, and parameter estimation was
carried out using a Bayesian approach.
\newline

\noindent\textbf{The skew normal distribution of \citeauthor{J012} (B-MSN)} \newline
\cite{J012} generalized the original skew normal distribution
to the class of (restricted) skew elliptical distributions.
In their parameterization,
the term $\bD$ used in the A-MSN distribution was removed,
resulting in an algebraically simpler form.
However, under this variant parameterization,
a change in scale will affect the skewness parameter.
The reader is referred to \citet{J017} for a discussion
on the effects of adopting this parameterization.
The skew normal member of this family, denoted by B-MSN,
has density
\begin{equation}
f(\by; \bmu, \bdelta, \bSigma) = 2 \phi_p(\by; \bmu, \bSigma)
    \Phi_1(\bdelta^T\bSigma^{-1}(\by-\bmu);
    0,1-\bdelta^T\bSigma^{-1}\bdelta).
\label{BrancoSN}
\end{equation}

\noindent
It follows that the conditioning-type stochastic representation
for $\bY$ is given by \linebreak
$\bY = \bmu + (\bY_1\mid Y_0 > 0)$, where
\begin{eqnarray}
\left[\begin{array}{c} Y_0 \\ \bY_1 \end{array}\right]
        \sim N_{1+p} \left(\left[\begin{array}{c} 0 \\
        \bzero \end{array}\right],
        \left[\begin{array}{cc} 1 & \bdelta^{T} \\
        \bdelta &
        \bSigma    \end{array}\right]\right),
\label{eq6}
\end{eqnarray}
and the corresponding convolution-type representation is
\begin{equation}
\bY = \bmu + \bdelta\left|\tilde{Y}_0\right| + \tilde{\bY}_1,
\label{eq7}
\end{equation}
where $\tilde{Y}_0 \sim N_1(0,1)$ and $\tilde{\bY}_1
\sim N_p(\bzero, \tilde{\bSigma})$ are independent,
and where $\tilde{\bSigma} = \bSigma - \bdelta\bdelta^T$.
Note that (\ref{eq6}) and (\ref{eq7}) are identical to 
(\ref{eq2.19}) and (\ref{eq2.20}) respectively.
It can be observed that (\ref{BrancoSN}) is a reparameterization
of the A-MSN distribution.
Replacing $\bdelta$ in (\ref{BrancoSN}) with $\bD\bdelta_A$
recovers (\ref{AzzaSN}).
\newline

\noindent\textbf{The skew normal/independent skew normal distribution (SNI-SN)}\newline
The \emph{skew normal Independent} (SNI) distributions are,
in essence, scale mixtures of the skew normal distribution.
Introduced by \citet{J012}, and considered further in \citet{J049},
the family includes the multivariate skew normal distribution
as the basic degenerate case, the density of which is given by
\begin{equation}
f(\by; \bmu, \bdelta_S, \bSigma) = 2 \phi_p(\by; \bmu, \bSigma)
    \Phi_1(\bdelta_S^T\bSigma^{-\frac{1}{2}}(\by-\bmu);
    0,1-\bdelta_S^T\bdelta_S),
\label{SNISN}
\end{equation}
where $\bSigma^{\frac{1}{2}}$ is the square root matrix of $\bSigma$;
that is, $\bSigma^{\frac{1}{2}} \bSigma^{\frac{1}{2}} = \bSigma$.
We shall adopt the notation $\bY \sim$ SNI-SN$_p(\bmu, \bSigma, \bdelta_S)$
when $\bY$ has density (\ref{SNISN}).
As with all restricted MSN distributions, the SNI-SN distribution
also enjoys two parallel stochastic representations.
This density is very similar to (\ref{AzzaSN}) and (\ref{BrancoSN}),
and actually, is a reparameterization of them.
The connection between them can be easily observed by directly comparing
their stochastic representations. The two stochastic representations
of the SNI-SN are given by
\begin{eqnarray}
\bY &=& \bmu + (\bY_1\mid Y_0>0),
\label{eq.2.21a}
\end{eqnarray}
and
\begin{eqnarray}
\bY &=& \bmu + \bSigma^{\frac{1}{2}}\bdelta_S|\tilde{Y}_0| 
	+ \bSigma^{\frac{1}{2}}(I_p-\bdelta_S\bdelta_S^T)^{\frac{1}{2}} 
	\tilde{\bY}_1,
\label{eq2.21b}        
\end{eqnarray}
where
\begin{eqnarray}
    {\left[\begin{array}{c} Y_0 \\ \bY_1 \end{array}\right]
        \sim N_{1+p} \left(\left[\begin{array}{c} 0 \\
        \bzero \end{array}\right],
        \left[\begin{array}{cc} 1
        & \bdelta_S^{T}\bSigma^{\frac{1}{2}} \\
        \bSigma^{\frac{1}{2}}\bdelta_S & \bSigma    
        \end{array}\right]\right),}  
\end{eqnarray}
and $\tilde{Y}_0 \sim N_1(0, 1), \; \tilde{\bY}_1 \sim N_p(\bzero, \bI_p)$
are independent.
It can be observed that (\ref{SNISN}) becomes identical
to (\ref{BrancoSN}) by replacing $\bdelta$ in (\ref{SNISN})
with $\bSigma^{\frac{1}{2}}\bdelta_S$.
\citet{J066} described maximum likelihood (ML) estimation for
the SNI-SN distribution via the expectation-maximization (EM) algorithm,
and an extension to the mixture model was also studied.
\newline

\noindent\textbf{The skew normal distribution of \citeauthor{J004} (P-MSN)}\newline
In a study of automated flow cytometry analysis,
\cite{J004} proposed yet another parametrization of
the restricted skew normal distribution.
This variant, hereafter referred to as the rMSN distribution
(as used in \citet{J103}), was obtained as
a `simplification' of the unrestricted skew normal distribution
described in \citet{J002} (see Section \ref{sec:uMSN}).
Its density is given by
\begin{equation}
f\left(\by; \bmu, \bSigma, \bdelta\right)
        = 2 \phi_p\left(\by; \boldsymbol \mu, \bSigma\right)
        \Phi_1\left(\bdelta^T \bSigma^{-1}
        \left(\by - \bmu\right);0,1-\bdelta^T\bSigma^{-1}\bdelta\right).
\label{SamSN}
\end{equation}
It follows that the conditioning-type stochastic representation
of (\ref{SamSN}) is given by
\begin{equation}
\bY = \bmu + \left(\bY_1 \mid Y_0 > 0\right),
\label{eq2.22}
\end{equation}
where
\begin{eqnarray}
\left[\begin{array}{c} Y_0 \\ \bY_1 \end{array}\right]
        \sim N_{1+p} \left(\left[\begin{array}{c} 0 \\
        \bzero \end{array}\right],
        \left[\begin{array}{cc} 1 & \bdelta^{T} \\
        \bdelta & \bSigma \end{array}\right]\right),
\label{eq10}
\end{eqnarray}
and the corresponding convolution-type representation is given by
\begin{equation}
\bY = \bmu + \bdelta|\tilde{Y}_0| + \tilde{\bY}_1,
\label{eq11}
\end{equation}
where again $\tilde{Y}_0 \sim N_1(0,1)$ and
$\tilde{\bY}_1 \sim N_p(\bzero, \tilde{\bSigma})$ are independent,
and where $\tilde{\bSigma} = \bSigma - \bdelta \bdelta^T$.
It can be observed that (\ref{SamSN}) is identical to (\ref{BrancoSN}).
One advantage of this parameterization is that the convolution-type
representation is in a relatively simple form,
and leads to a nice hierarchical form
which facilitates implementation of the EM algorithm
for ML parameter estimation.   

For ease of reference, we include a summary of the density
and stochastic representation of the above-mentioned
restricted MSN distributions in Table \ref{tab:rMSNa}
and \ref{tab:rMSNb}, respectively.
\newline

\begin{table}
    \centering
        \begin{tabular}{|c|c|}
            \hline
            Distribution & Density \\
            \hline
                &    \\    
            {A-MSN}    &        
                $f\left(\by\right) =
                2 \phi_p\left(\by; \bmu, \bSigma\right)
                \Phi_1\left(\bdelta_A^T \bR^{-1} \bD^{-1}
                \left(\by - \bmu\right);
                0, 1-\bdelta_A^T\bR^{-1}\bdelta_A\right)$    \\
            (1996)    &
            $\bD=\mbox{diag}(\sqrt{\Sigma_{11}},\cdots,\sqrt{\Sigma_{pp}}),
            \;\;  \bR = \bD^{-1}\bSigma\bD^{-1}$    \\
                &    \\    
            \hline    
                    
                &    \\    
            {B-MSN}    &                
            \multirow{2}{*} {$f(\by) = 2 \phi_p(\by; \bmu, \bSigma)
            \Phi_1(\bdelta^T\bSigma^{-1}(\by-\bmu);
            0,1-\bdelta^T\bSigma^{-1}\bdelta)$}
             \\
            (2001) & \\
                &    \\    
            \hline
            
                &    \\    
            {P-MSN}    &    
            \multirow{2}{*} {$f(\by) = 2 \phi_p(\by; \bmu, \bSigma)
            \Phi_1(\bdelta^T\bSigma^{-1}(\by-\bmu);
            0,1-\bdelta^T\bSigma^{-1}\bdelta)$} 
             \\
            (2009) &        \\
                &    \\    
            \hline
            
                &    \\    
            {SNI-SN} &     
            \multirow{2}{*} {$f(\by)
            = 2 \phi_p(\by; \bmu, \bSigma) \Phi_1\left(\bdelta_S^T
            \bSigma^{-\frac{1}{2}} \left(\by - \bmu\right);
            0,1-\bdelta_S^T\bdelta_S\right)$} \\
            (2010)&     \\
                &    \\    
            \hline
        \end{tabular}
    \caption{Summary of the densities of selected restricted forms
    of multivariate skew normal distributions.}
    \label{tab:rMSNa}
\end{table}

\begin{landscape}
\begin{table}
    \centering
        \begin{tabular}{|c|c|c|}
            \hline
            Distribution    &    Conditioning-type representation
                & Convolution-type representation \\
            \hline
                &	&    \\    
            {A-MSN}    &        
                $\bY = \bmu + \bD (\bY_1\mid Y_0 > 0)$  &    
                $\bY = \bmu + \bD\bdelta_A|\tilde{Y}_0|+\tilde{\bY}_1$\\
            (1996)    &
                $\left[\begin{array}{c}Y_0\\\bY_1\end{array}\right]
                \sim N_{1+p}\left(         
                \left[\begin{array}{c}0\\\bzero\end{array}\right],
                \left[\begin{array}{cc}1&\bdelta_A^T\     
                \\\bdelta_A&\bR\end{array}\right]\right)$    &    
                $\begin{array}{rcl} \tilde{Y}_0 &\sim& N_1(0, 1) \\
                \tilde{\bY}_1 & \sim & 
                N_p(\bzero, \bSigma-\bD\bdelta_A\bdelta_A^T\bD)
                \end{array}$    \\            
                &   &	 \\    
            \hline
            
                &  &  \\    
            {B-MSN}    &    
                $\bY = \bmu + (\bY_1\mid Y_0>0)$ &    \
                $\bY = \bmu + \bdelta |\tilde{Y}_0| + \tilde{\bY}_1$\\
            (2001) & $\left[\begin{array}{c}Y_0\\\bY_1\end{array}\right]
                \sim N_{1+p}\left(     
                \left[\begin{array}{c}0\\\bzero\end{array}\right],
                \left[\begin{array}{cc}1&\bdelta^T\     
                \\\bdelta&\bSigma\end{array}\right]\right)$ &
                $\begin{array}{rcl} \tilde{Y}_0 &\sim& N_1(0,1) \\
                \tilde{\bY}_1 &\sim& N_p(\bzero, \bSigma-\bdelta\bdelta^T)
                \end{array}$    \\
                &   &	 \\    
            \hline
 
                 &  &  \\    
            {P-MSN}    &    
                $\bY = \bmu + (\bY_1 \mid Y_0 > 0)$    &
                $\bY = \bmu + \bdelta|\tilde{Y}_0| + \tilde{\bY}_1$ \\
            (2009) &
                $\left[\begin{array}{c} Y_0\\{\bY}_1\end{array}\right]
                \sim N_{1+p}\left(     
                \left[\begin{array}{c} 0 \\\bzero\end{array}\right],
                \left[\begin{array}{cc} 1 & \bdelta^T     
                \\\bdelta &\bSigma\end{array}\right]\right)$    &    
                $\begin{array}{rcl} \tilde{Y}_0 &\sim& N_1(0, 1) \\
                \tilde{\bY}_1 & \sim & N_p(\bzero, \tilde{\bSigma}) \\
                \tilde{\bSigma} &=& \bSigma - \bdelta\bdelta^T
                \end{array}$ \\
                &	&    \\    
            \hline

                &  &	  \\    
            {SNI-SN} &         
                $\bY = \bmu + (\bY_1 \mid Y_0 > 0)$  &
                $\bY = \bmu + \bSigma^{\frac{1}{2}}\bdelta_S|\tilde{Y}_0| 
                + \bSigma^{\frac{1}{2}} (I_p-\bdelta_S\bdelta_S^T)
                ^{\frac{1}{2}} \tilde{\bY}_1$ \\
            (2010)&    
                $\left[\begin{array}{c} Y_0\\{\bY}_1\end{array}\right]
                \sim N_{1+p}\left(     
                \left[\begin{array}{c}0\\\bzero\end{array}\right],
                \left[\begin{array}{cc}1&\bdelta_S^T\bSigma^{\frac{1}{2}}     
                \\\bSigma^{\frac{1}{2}}\bdelta_S&\bSigma\end{array}\right]\right)$ &
                $\begin{array}{rcl} \tilde{Y}_0 &\sim& N_1(0, 1) \\
                \tilde{\bY}_1 & \sim & N_p(\bzero, \bI_p)
                \end{array}$ \\
                &	&    \\    
            \hline
        \end{tabular}
    \caption{Summary of stochastic representations of
    selected restricted forms of multivariate skew normal distributions.}
    \label{tab:rMSNb}
\end{table}
\end{landscape}

\subsubsection{Unrestricted multivariate skew normal distributions}
\label{sec:uMSN}

The unrestricted case is very similar to the restricted case,
except that the scalar latent variable is replaced by
a $p$-dimensional normal random vector $\bY_0$.
Accordingly, the constraint $Y_0 > 0$ becomes a set of
$p$ constraints $\bY_0 > \bzero$, which implies each element
of $\bY_0$ is positive. Similar to (\ref{eq2.18}) and (\ref{eq2.19}),
the unrestricted MSN (uMSN) distribution can be described by
\begin{equation}
\bY = \bmu + (\bY_1 \mid \bY_0 > \bzero),
\label{eq12}
\end{equation}
where
\begin{equation}
\left[\begin{array}{c} \bY_0 \\ \bY_1 \end{array}\right]
        \sim N_{2p} \left(\left[\begin{array}{c} \bzero \\
        \bzero \end{array}\right],
        \left[\begin{array}{cc} \bI_p & \bDelta^T \\
        \bDelta & \bSigma
        \end{array}\right]\right).
\label{eq13}
\end{equation}
Here, the skewness parameter $\bDelta$ is a $p\times p$ matrix.
The convolution-type representation is analogous to (\ref{eq2.20}),
and is given by
\begin{equation}
\bY = \bmu + \tilde{\bDelta}|\tilde{\bY}_0|
    + \tilde{\bY}_1,
\label{eq14}
\end{equation}  
where the random vectors $\tilde{\bY}_0 \sim N_p(\bzero, \bI_p)$
and $\tilde{\bY}_1 \sim N_p(\bzero, \tilde{\bSigma})$ 
are independently distributed.
The relationship between the parameters 
in (\ref{eq13}) and (\ref{eq14}) is similar to that 
in (\ref{eq2.18})-(\ref{eq2.20}).
In this case, they satisfy 
$\tilde{\bDelta} = \bDelta$ and
$\tilde{\bSigma} = \bSigma - \bDelta\bDelta^{T}$. 
The skew normal version of \citet{J002} is
an unrestricted form of the MSN distribution,
with $\bDelta$ restricted to be a diagonal matrix.
\newline

\noindent\textbf{The skew normal distribution of \citeauthor{J002} (S-MSN)}\newline
In \cite{J002}, skewness is introduced to a class of
elliptically symmetric distributions by conditioning
on a multivariate variable, which produces
a class of (unrestricted) skew elliptical distribution.
The multivariate skew normal distribution
proposed by \citet{J002}, which is a member of this family,
is given by
\begin{equation}
f\left(\by; \bmu, \bSigma, \bdelta\right)
     = 2^p \phi_p\left(\by; \bmu, \bSigma\right)
     \Phi_p\left(\bDelta \bSigma^{-1}\left(\by - \bmu\right);
     \bzero, \bLambda\right),
\label{SahuSN}
\end{equation}
where $\bDelta = \mbox{diag} \left(\bdelta\right)$
and $\bLambda = \bI_p - \bDelta \bSigma^{-1} \bDelta$.
Observe that with this characterization of the MSN distribution,
the density involves the \emph{multivariate}
normal distribution function, whereas the restricted forms
is defined in terms of the \emph{univariate} distribution instead.
Accordingly, the conditioning-type stochastic representation
of (\ref{SahuSN}) is given by 
$\bY = \bmu + \left(\bY_1 \mid \bY_0 > \bzero\right)$,
where
\begin{eqnarray}
\left[\begin{array}{c} \bY_0 \\ \bY_1 \end{array}\right]
        \sim N_{2p} \left(\left[\begin{array}{c} \bzero \\
        \bzero \end{array}\right],
        \left[\begin{array}{cc} \bI_p & \bDelta \\
        \bDelta & \bSigma
        \end{array}\right]\right),
\label{Eq15}
\end{eqnarray}
and the convolution-type representation is given by
\begin{equation}
\bY = \bmu + \bDelta |\tilde{\bY}_0| + \tilde{\bY}_1,
\label{eq16}
\end{equation}
where $\tilde{\bY}_0$ and $\tilde{\bY}_1$ are independent variables
distributed as $\tilde{\bY}_0 \sim N_p(\bzero, \bI_p)$
and $\tilde{\bY}_1 \sim N_p(\bzero, \tilde{\bSigma})$, respectively,
and where $\tilde{\bSigma} = \bSigma - \bDelta^2$.
ML estimation for the uMSN distribution, and its mixture case,
is studied in \citet{J048}.
\newline

\subsubsection{Extended multivariate skew normal distributions}
\label{sec:eMSN}

We consider now the extended skew normal (ESN) distribution,
which originates from a selective sampling problem,
where the variable of interest is affected by
a latent variable that is truncated at an arbitrary threshold.
It can be obtained via conditioning by setting
$\bY = \bmu + (\bY_1\mid Y_0 + \tau > 0)$,
where $\bY_1$ and $Y_0$ are distributed according to (\ref{eq2.19}),
which leads to the density
\begin{equation}
f(\by; \bmu, \bSigma, \tau)
    = \phi_p(\by; \bmu, \bSigma)
    \frac{\Phi_1\left(\tau + \bdelta^T\bSigma^{-1}(\by-\bu);
    0, 1-\bdelta^T\bSigma^{-1}\bdelta\right)}
    {\Phi_1(\tau;0,1)}.
\label{ESN}
\end{equation}
This expression for an ESN distribution is due to \citet{J022},
and the threshold $\tau$ is known as an extension parameter.
With this additional parameter, the normalizing constant
is no longer a simple fixed value
(such as $2$ in the restricted case and $2^p$ in the unrestricted case),
but a scalar value that depends on the extension parameter.  
Although the ESN is more complicated than the restricted
and unrestricted skew normal distributions,
it has nice properties not shared by these `no-extension' cases,
including closure under conditioning.

The ESN distribution represents one of the simplest cases
of the extended form. Replacing the latent variable $Y_0$
with a $q$-dimensional version $\bY_0$ leads to
the unified skew normal (SUN) distribution \citep{J017}.
The SUN distribution is an attempt to unify all of
the aforementioned skew normal distributions.
Its conditioning-type stochastic representation
is given by (\ref{eq2.16}) and (\ref{eq2.17}).
It follows that the SUN density is given by
\begin{equation}
f(\by; \bmu, \bSigma, \bGamma, \bDelta, \btau)
    = \phi_p(\by; \bmu, \bSigma)
    \frac{\Phi_q\left(\btau + \bDelta^T\bSigma^{-1}(\by-\bmu);
    \bzero, \bGamma - \bDelta^T\bSigma^{-1}\bDelta\right)}
    {\Phi_q(\btau; \bzero, \bGamma)}.
\label{SUN}
\end{equation}
 
Its construction can also be achieved via the convolution approach,
where the $q$-dimensional latent variable $\bY_0$ follows
a truncated normal distribution with mean $\btau$.
More specifically, let $\tilde{\bY}_1 \sim N_p(0, \bSigma)$
and $\tilde{\bY}_0 \sim TN_q(\btau, \bGamma)$ be independent variables,
where $TN_q(\btau, \bGamma)$ denotes a multivariate normal variable
with mean vector $\btau$ and covariance $\bGamma$ truncated to
the positive hyperplane.
Then $\bY = \bmu + \bDelta \tilde{\bY}_0 + \tilde{\bY}_1$ 
has an extended MSN density.
Note that in this case, the skewness parameter is
a $p\times q$ matrix instead of the $p$-dimensional vector
$\bdelta$ used in the restricted and unrestricted forms
of the MSN distribution.

It is not difficult to show that the SUN distribution
includes the restricted MSN distributions,
the unrestricted MSN distributions,
and the ESN distribution as special cases.
There are also various versions of MSN distributions
which turns out to be equivalent to the SUN distribution,
including the hierarchical skew normal (HSN) of \citet{J015},
the closed skew normal (CSN) of \citet{J018},
the skew normal of \citet{J014}
and a location-scale variant of the 
canonical fundamental skew normal (CFUSN) distribution \citep{J008}.
For a detail discussion on the equivalence between
these extended forms of MSN distributions,
the reader is referred to \citet{J017}.

\subsubsection{Generalized multivariate skew normal distributions}
\label{sec:gMSN}

A further generalization of the extended form of the MSN distribution
is to relax the distributional assumption of the latent variable $\bY_0$.  
For the `generalized form' of the MSN distribution,
there are no other restrictions on the MSN density
except that the symmetric part must be a multivariate normal density,
that is, $\bY_1$ is normally distributed.
This form is very general and apparently
includes the other three forms discussed above.
A prominent example is the
\emph{fundamental skew normal distribution} (FUSN),
a member of the class of fundamental skew distributions
considered by \citet{J008}. Its density is given by
\begin{equation}
f(\by; \bmu, \bSigma, Q_q) = K_q^{-1}
    \phi_p(\by; \bmu, \bSigma) Q_q(\by),
\label{SN3a}
\end{equation}
where $K_q = E\left\{Q_q(\bY)\right\}$ is a normalizing constant
and $Q_q(\by)$ is a \emph{skewing function}.
Notice that the skewing function here
is not restricted to the normal family.
As mentioned previously, the FUSN density can be obtained by
defining $\bY = \left(\bY_1 \mid \bY_0 > \bzero\right)$,
where $\bY_1$ follows the $p$-dimensional normal distribution
with location parameter $\bmu$ and scale matrix $\bSigma$
and $\bY_0$ is a $q\times 1$ random vector.
Under this definition, $K_q$ and $Q_q(\by)$
is given by $P(\bY_0>\bzero)$ and \linebreak
$P(\bY_0>\bzero \mid \bY_1)$, respectively.   

An interesting special case of (\ref{SN3a}) is
the location-scale variant of the so-called
\emph{canonical fundamental skew normal} (CFUSN) distribution,
obtained by taking $\bY_0 \sim N_q(\bzero, \bI_q)$
and cov$(\bY_1, \bY_0) = \bDelta$.
In this case, we have
$\bY_0\mid\bY_1 \sim N_q(\bDelta^T\bSigma^{-1}(\by-\bmu), \bLambda)$,
where $\bLambda=\bI_q - \bDelta^T\bSigma^{-1}\bDelta$.
This leads to the density
\begin{equation}
f(\by;\bmu, \bSigma, \bDelta) = 2^q \phi_p(\by; \bmu, \bSigma) 
	\Phi_q(\bDelta^T\bSigma^{-1}(\by-\bmu); \bzero, \bLambda).
\label{SN3}
\end{equation}
We shall write $\bY \sim CFUSN_{p,q}(\bmu, \bSigma, \bDelta)$.
It should be noted that by taking  $q=p$
and $\bDelta = \mbox{diag}(\bdelta)$,
(\ref{SN3}) reduces to
the unrestricted skew normal density introduced by \cite{J002}.
Also, the CFUSN density reduces to
the restricted B-MSN distribution (\ref{BrancoSN})
when $q=1$ and $\bDelta = \bdelta$.

\subsection{Multivariate skew $t$-distributions}
\label{sec:MST}

The multivariate skew $t$-distribution is an important member
of the family of skew-elliptical distributions.
Like the skew normal distributions,
there exists various different versions of the MST distribution,
which can be na\.ively classified into four broad forms. 
The MST distribution is of special interest because
it offers greater flexibility than the normal distribution
by combining both skewness and kurtosis in its formulation,
while retaining a fair degree of tractability in an algebraic sense.
This additional flexibility is much needed in some practical applications,
as will be demonstrated in the example in Section \ref{sec:DLBCL}.

In the past two decades, many variants of the
multivariate skew $t$-distribution have been proposed.
Some notable proposals include
the skew $t$-member of \citet{J012}'s skew elliptical class,
the skew $t$-distribution of \citet{J006},
the skew $t$-distribution of \citet{J019},
the skew $t$-distribution of \citet{J002}'s skew elliptical class,
the skew normal/independent skew $t$ (SNI-ST) distribution of \citet{J049},
the closed skew $t$ (CST) distribution of \citet{T001},
and the extended skew $t$ (EST) distribution of \citet{J074}.
Many of these can be considered as special cases of
the fundamental skew $t$ (FUST) distribution
introduced by \citet{J008}.
They may be classified as `restricted', `unrestricted',
`extended', and `generalized' subclasses of the FUST distribution
(see Table \ref{tab:2.4}).

\begin{table}
    \centering
        \begin{tabular}{|c|c|c|}
            \hline
            Case    &    Restrictions on FUST    & Examples    \\
            \hline
            restricted    &    $q=1$, $\tau=0$ and
                $\left[\begin{array}{c}Y_0 \\ \bY_1\end{array}\right]
                \sim t_{1+p}$  &    
            B-MST, A-MST, G-MST, P-MST, SNI-ST\\
            \hline
            unrestricted    &    $q=p$,
                $\left[\begin{array}{c} \bY_0 \\ \bY_1\end{array}\right]
                \sim t_{2p}$ &
            S-MST\\
            \hline
            extended    &    
                $\left[\begin{array}{c} \bY_0 \\ \bY_1\end{array}\right]
                \sim t_{q+p}$ &
                EST, CST, CFUST, SUT    \\
            \hline
            generalized    &    $\bY_1$ is $t$-distributed  &    FST, GST \\
            \hline
        \end{tabular}
    \caption{Classification of MST distributions.}
    \label{tab:2.4}
\end{table}

\begin{table}
    \centering
    \begin{tabular}{|c|c|c|}
        \hline
        Abbreviation    &    Name    &    References    \\        
        \hline
        \multicolumn{1}{|c}{\textbf{rMST}}    & \multicolumn{1}{c}{}     &    \\
        \hline
        B-MST    &    Branco's MST    	&    \citet{J012}    \\
        A-MST    &    Azzalini's MST    &    \citet{J006}    \\
        G-MST    &    Gupta's MST       &    \citet{J019}    \\
        P-MST    &    Pyne's MST        &    \citet{J004}    \\    
        SNI-ST   &    skew normal/independent    MST&    \citet{J049} \\
        \hline
        \multicolumn{1}{|c}{\textbf{uMST}}    &    \multicolumn{1}{c}{} &    \\
        \hline
        S-MST    &    Sahu's MST    	&    \citet{J002}    \\
        \hline
        \multicolumn{1}{|c}{\textbf{eMST}}    & \multicolumn{1}{c}{}     &    \\
        \hline
        EST    &    Extended MST    	&    \citet{J074}    \\
        CST    &    Closed    MST       &    \citet{T001}    \\
        SUT    &    Unified MST         &    \citet{J017} \\
        \hline        
        \multicolumn{1}{|c}{\textbf{gMST}}    & \multicolumn{1}{c}{}     &    \\
        \hline
        FUST    &    Fundamental MST    &    \citet{J008}    \\
        GST     &    Generalized MST    &    \citet{J023}    \\
        FST     &    Flexible MST    	&    \citet{J072}    \\
        \hline                    
    \end{tabular}
    \caption{Summary of the abbreviations of skew $t$-distributions
    used in Table \ref{tab:2.4}.}
    \label{tab2.5}
\end{table}

\subsubsection{Restricted multivariate skew $t$-distributions}
\label{sec:rMST}

The restricted skew $t$-distribution is obtained by conditioning on
a univariate latent variable $Y_0$ being positive.
The correlation between $\bY_1$ and $Y_0$
is described by the vector $\bdelta$.
Like the MSN distributions, the MST distributions
can be obtained via a conditioning and convolution mechanism.
In general, the restricted MST distribution
has a conditioning-type stochastic representation given by:
\begin{equation}
\bY = \bmu + (\bY_1\mid Y_0 > 0),
\label{eq2.23}
\end{equation}
where
\begin{equation}
\left[\begin{array}{c} Y_0 \\ \bY_1 \end{array}\right]
        \sim t_{1+p} \left(\left[\begin{array}{c} 0 \\
        \bzero \end{array}\right],
        \left[\begin{array}{cc} 1 & \bdelta^T \\
        \bdelta & \bSigma
        \end{array}\right],\nu \right).
\label{eq2.23b}
\end{equation}
The equivalent convolution-type representation is given by
\begin{equation}
\bY = \bmu + \tilde{\bdelta}|\tilde{Y}_0| + \tilde{\bY}_1,
\label{eq2.24}
\end{equation}
where the two random variables $\tilde{Y}_0$ and $\tilde{\bY}_1$ 
have a joint multivariate central $t$-distribution with scale matrix
$\left[\begin{array}{cc} 1 & \bzero \\ \bzero 
& \tilde{\bSigma} \end{array}\right]$ 
and $\nu$ degrees of freedom.
The link between the pairs of parameters $(\bdelta,\bSigma)$
and $(\tilde{\bdelta},\tilde{\bSigma})$ is the same as that
for the rMSN distribution.
The skew-$t$ distribution of \citet{J012}, \citet{J006},
\citet{J019}, the SNI-ST, and the skew $t$-version given by \citet{J004}
are equivalent to (\ref{eq2.23}) up to a reparametrization.
\newline

\begin{table}
    \centering
        \begin{tabular}{|c|c|}
            \hline
            Name & Density    \\
            \hline
            	&	\\	
            {B-MST}    &    
            $f(\by) = 2 t_p(\by; \bmu, \bSigma, \nu)
            T_1(\bdelta^T\bSigma^{-1}(\by-\bmu)
            \sqrt{\frac{\nu+p}{\nu+d(\by)}};
            0,1-\bdelta^T\bSigma^{-1}\bdelta, \nu+p)$ \\
            (2001) &    $d(\by) = (\by-\bmu)^T \bSigma^{-1} (\by-\bmu)$     \\
            	&	\\	
            \hline
            
            	&	\\	
            &        
            $f\left(\by\right) = 2 t_p\left(\by; \bmu, \bSigma, \nu \right)
            T_1\left(\bdelta_A^T \bR^{-1} \bD^{-1} \left(\by - \bmu\right)
            \sqrt{\frac{\nu+p}{\nu+d_(\by)}};
            0, 1-\bdelta_A^T\bR^{-1}\bdelta_A, \nu+p\right)$ \\
            {A-MST}    &
            $\bD=\mbox{diag}(\sqrt{\Sigma_{11}},\cdots,\sqrt{\Sigma_{pp}}),$ \\
            (2003)    & $\bR = \bD^{-1}\bSigma\bD^{-1}$      \\
            &    $d(\by) = (\by-\bmu)^T \bSigma^{-1}(\by-\bmu)$    \\
            	&	\\	
            \hline    
                    
            	&	\\	
            {G-MST}    &    
            $f(\by) = 2 t_p(\by; \bmu, \bSigma, \nu)
            T_1(\bdelta_G^T(\by-\bmu)    \sqrt{\frac{\nu+p}{\nu+d(\by)}};
            0,1-\bdelta_G^T\bSigma\bdelta_G, \nu+p)$ \\
            (2003) &    $d(\by) = (\by-\bmu)^T \bSigma^{-1} (\by-\bmu)$ \\
            	&	\\	
            \hline
            
            	&	\\	
            {P-MST}    &    
            $f(\by) = 2 t_p(\by; \bmu, \bSigma, \nu) 
            T_1\left(\bdelta^T \bSigma^{-1}
            \left(\by - \bmu\right) \sqrt{\frac{\nu+p}{\nu+d(\by)}};
            0,1-\bdelta^T\bSigma^{-1}\bdelta, \nu+p\right)$ \\
            (2009) &  $d(\by) = (\by-\bmu)^T \bSigma^{-1} (\by-\bmu)$  \\
            	&	\\	
            \hline
            
            	&	\\	
            {SNI-ST} &     
            $f(\by) = 2 t_p(\by; \bmu, \bSigma, \nu) T_1\left(\bdelta_S^T
            \bSigma^{-\frac{1}{2}} \left(\by - \bmu\right)
            \sqrt{\frac{\nu+p}{\nu+d(\by)}};
            0,1-\bdelta_S^T\bdelta_S, \nu+p\right)$  \\
            (2010)&    $d(\by) = (\by-\bmu)^T \bSigma^{-1} (\by-\bmu)$    \\
            	&	\\	
            \hline
        \end{tabular}
    \caption{Densities of selected restricted forms of multivariate
    skew $t$-distributions.}
    \label{tab:rMSTa}
\end{table}

\begin{landscape}
\begin{table}
    \centering
        \begin{tabular}{|c|c|c|}
            \hline
            Distribution    &    Conditioning-type representation
                & Convolution-type representation \\            
            \hline
            	&	&	\\	
            {B-MST}    &    
                $\bY = \bmu + (\bY_1\mid Y_0>0)$    &
                $\bY = \bmu + \bdelta|\tilde{Y}_0| + \tilde{\bY}_1$ \\
            (2001)
                & $\left[\begin{array}{c}Y_0\\\bY_1\end{array}\right]
                \sim t_{1+p}\left(     
                \left[\begin{array}{c}0\\\bzero\end{array}\right],
                \left[\begin{array}{cc}1&\bdelta^T\     
                \\\bdelta&\bSigma\end{array}\right], \nu\right)$    &
                $\left[\begin{array}{c} \tilde{Y}_0\\\tilde{\bY}_1
                \end{array}\right]
                \sim t_{1+p}\left(     
                \left[\begin{array}{c}0\\\bzero\end{array}\right],
                \left[\begin{array}{cc}1&\bzero 
                \\\bzero&\bSigma-\bdelta\bdelta^T\end{array}\right], 
                \nu\right)$    \\
            	&	&	\\	
            \hline
            
            	&	&	\\	
            {A-MST}    &        
                $\bY = \bmu + \bD (\bY_1\mid Y_0 > 0)$  &    
                $\bY = \bmu + \bD\bdelta_A |\tilde{Y}_0| + \tilde{\bY}_1$    \\
            (2003)    &
                $\left[\begin{array}{c}Y_0\\\bY_1\end{array}\right]
                \sim t_{1+p}\left(         
                \left[\begin{array}{c}0\\\bzero\end{array}\right],
                \left[\begin{array}{cc}1&\bdelta_A^T\     
                \\\bdelta_A&\bR\end{array}\right], \nu\right)$    &
                $\left[\begin{array}{c} \tilde{Y}_0\\\tilde{\bY}_1
                \end{array}\right]
                \sim t_{1+p}\left(     
                \left[\begin{array}{c}0\\\bzero\end{array}\right],
                \left[\begin{array}{cc}1&\bzero 
                \\\bzero&\bSigma - \bD\bDelta_A\bdelta_A^T\bD
                \end{array}\right], \nu\right)$    \\
            	&	&	\\	
            \hline    
                    
            	&	&	\\	
            {G-MST}    &    
                $\bY = \bmu + (\bY_1\mid Y_0>0)$ &    \
                $\bY = \bmu + \bSigma\bdelta_G |\tilde{Y}_0| + \tilde{\bY}_1$ \\
            (2003)
                & $\left[\begin{array}{c}Y_0\\\bY_1\end{array}\right]
                \sim t_{1+p}\left(     
                \left[\begin{array}{c}0\\\bzero\end{array}\right],
                \left[\begin{array}{cc}
                1 &\bdelta_G^T\bSigma     
                \\\bSigma\bdelta_G&\bSigma\end{array}\right], \nu\right)$    &
                $\left[\begin{array}{c} \tilde{Y}_0\\\tilde{\bY}_1
                \end{array}\right]
                \sim t_{1+p}\left(     
                \left[\begin{array}{c}0\\\bzero\end{array}\right],
                \left[\begin{array}{cc}1&\bzero 
                \\\bzero&\bSigma-\bSigma\bdelta_G\bdelta_G^T\bSigma
                \end{array}\right], \nu\right)$    \\
            	&	&	\\	
            \hline
            
            	&	&	\\	
            {P-MST}    &
                $\bY = \bmu + (\bY_1 \mid Y_0 > 0)$    &
                $\bY = \bmu + \bdelta|\tilde{Y_0}| + \tilde{\bY}_1$ \\
            (2009) &    
                $\left[\begin{array}{c}{Y}_0 \\
                \bY_1\end{array}\right]
                \sim t_{1+p}\left(
                \left[\begin{array}{c} 0 \\\bzero\end{array}\right],
                \left[\begin{array}{cc}1    &    \bdelta^T    
                \\\bdelta&\bSigma\end{array}\right], \nu\right)$ &
                $\begin{array}{c}
                \left[\begin{array}{c} \tilde{Y}_0\\\tilde{\bY}_1
                \end{array}\right]
                \sim t_{1+p}\left(     
                \left[\begin{array}{c}0\\\bzero\end{array}\right],
                \left[\begin{array}{cc}1&\bzero 
                \\\bzero&\tilde{\bSigma}\end{array}\right], \nu\right) \\
                \tilde{\bSigma} =\bSigma-\bdelta\bdelta^T
                \end{array}$\\
            	&	&	\\	
            \hline
            
            	&	&	\\	
            {SNI-ST} &     
                $\bY = \bmu + (\bY_1 \mid Y_0 > 0)$    &
                $\bY = \bmu + \bSigma^{\frac{1}{2}}\bdelta_S|\tilde{Y}_0| +
                \bSigma^{\frac{1}{2}}(I_p-\bdelta_S\bdelta_S^T)
                ^{\frac{1}{2}}\tilde{\bY}_1$  \\
            (2010)&            
            $\left[\begin{array}{c} {Y}_0\\{\bY}_1\end{array}\right]
            \sim t_{1+p}\left(     
            \left[\begin{array}{c} 0 \\\bzero\end{array}\right],
            \left[\begin{array}{cc}1 & \bdelta_S^T\bSigma^{\frac{1}{2}}     
                \\\bSigma^{\frac{1}{2}}\bdelta_S &\bSigma\end{array}\right], 
                \nu\right)$ &    
           $\left[\begin{array}{c} \tilde{Y}_0\\\tilde{\bY}_1
                \end{array}\right]
                \sim t_{1+p}\left(     
                \left[\begin{array}{c}0\\\bzero\end{array}\right],
                \left[\begin{array}{cc}1&\bzero 
                \\\bzero&\bI_p\end{array}\right], \nu\right)$ \\
            	&	&	\\	
            \hline
        \end{tabular}
    \caption{Stochastic representations of selected restricted forms 
    	of multivariate skew $t$-distributions.}
    \label{tab:rMSTb}
\end{table}
\end{landscape}

\noindent\textbf{The skew \emph{t}-distribution of \citeauthor{J012} (B-MST)}\newline
The skew elliptical class of \citet{J012} includes
a skew $t$-distribution, which is a special case of
a scale mixture of the B-MSN distribution.
Its density is given by
\begin{eqnarray}
f(\by) &=& 2 t_p(\by; \bmu, \bSigma, \nu) \; \nonumber\\
    &&        T_1\left(\bdelta^T\bSigma^{-1}(\by-\bmu)
    		\sqrt{\frac{\nu+p}{\nu+d(\by)}};
            0, 1-\bdelta^T\bSigma^{-1}\bdelta,\nu+p\right),
\label{BrancoST}
\end{eqnarray}
where $d(\by) = (\by-\bmu)^T\bSigma^{-1}(\by-\bmu)$
is the squared Mahalanobis distance between $\by$ and $\bmu$
with respect to $\bSigma$.
Here, we let $t_p(.;\bmu, \bSigma, \nu)$
denote the $p$-dimensional $t$-density with location vector $\bmu$,
scale matrix $\bSigma$, and degrees of freedom $\nu$,  
and $T_1(.;\mu, \sigma^2, \nu)$ denote the distribution function
of the (univariate) $t$-distribution with mean $\mu$,
variance $\sigma^2$ and degrees of freedom $\nu$.
It can  be observed from representation (\ref{BrancoST})
that the multivariate skew $t$-distribution converges to
the B-MSN density (\ref{BrancoSN})
when the degrees of freedom $\nu$ approaches infinity.

It follows that $\bY$ has a conditioning-type representation
given by $\bY = \bmu + (\bY_1\mid Y_0 > 0)$, where
\begin{eqnarray}
\left[\begin{array}{c} Y_0 \\ \bY_1 \end{array}\right]
        \sim t_{1+p} \left(\left[\begin{array}{c} 0 \\
        \bzero \end{array}\right],
        \left[\begin{array}{cc} 1 & \bdelta^{T} \\
        \bdelta &
        \bSigma    \end{array}\right], \nu \right),
\label{eq2.25}
\end{eqnarray}
and a corresponding convolution-type representation given by
\begin{equation}
\bY = \bmu + \bdelta\left|\tilde{Y}_0\right|
     + \tilde{\bY}_1,
\label{eq2.26}
\end{equation}
where 
\begin{eqnarray}
\left[\begin{array}{c} \tilde{Y}_0\\\tilde{\bY}_1
                \end{array}\right]
                \sim t_{1+p}\left(     
                \left[\begin{array}{c}0\\\bzero\end{array}\right],
                \left[\begin{array}{cc}1&\bzero 
                \\\bzero&\bSigma-\bdelta\bdelta^T\end{array}\right], \nu\right).
\nonumber
\end{eqnarray}
It can be seen that (\ref{eq2.25}) is identical
to (\ref{eq2.23b}).
\newline

\noindent\textbf{The skew \emph{t}-distribution of \citeauthor{J006} (A-MST)}\newline
\citet{J006} extended the A-MSN distribution of \citet{J001}
to the skew $t$-case. Its density is given by
\begin{eqnarray}
f\left(\by\right) &=& 2 t_p\left(\by; \bmu, \bSigma,\nu\right) \nonumber\\
    &    &    T_1\left(\bdelta_A^T \bR^{-1} \bD^{-1} \left(\by - \bmu\right)
    			\sqrt\frac{\nu+p}{\nu+d(\by)};
                0, 1-\bdelta_A^T\bR^{-1}\bdelta_A,\nu+p\right),
    \nonumber\\
\label{AzzaST}
\end{eqnarray}
where $d(\by) = (\by-\bmu)^T\bSigma^{-1}(\by-\bmu)$,
$\bR = \bD^{-1}\bSigma\bD^{-1}$ is the correlation matrix,
and $\bD$ is the diagonal matrix created by extracting
the diagonal elements of $\bSigma$.
Note again that the parameter $\bdelta$ in (\ref{AzzaST})
is marked with a subscript $A$ to indicate that it is different
to the definition used in (\ref{BrancoST})
and other rMST distributions.
The A-MST density (\ref{AzzaST}) can be obtained by
a conditioning mechanism, similar to the A-MSN distribution,
by setting \linebreak
$\bY = \bmu + \bD (\bY_1\mid Y_0 > 0)$, where
\begin{eqnarray}
\left[\begin{array}{c} Y_0 \\ \bY_1 \end{array}\right]
        \sim t_{1+p} \left(\left[\begin{array}{c} 0 \\
        \bzero \end{array}\right],
        \left[\begin{array}{cc} 1 & \bdelta_A^{T} \\
        \bdelta_A & \bR    
        \end{array}\right], \nu\right).
\label{eq2.27}
\end{eqnarray}
A parallel representation of (\ref{AzzaST})
via the convolution mechanism is given by
\begin{equation}
\bY = \bmu + \bD\bdelta_A\left|\tilde{Y}_0\right| + \tilde{\bY}_1,
\label{eq2.28}
\end{equation}
where 
\begin{eqnarray}
\left[\begin{array}{c} \tilde{Y}_0\\\tilde{\bY}_1
                \end{array}\right]
                \sim t_{1+p}\left(     
                \left[\begin{array}{c}0\\\bzero\end{array}\right],
                \left[\begin{array}{cc}1&\bzero 
                \\\bzero&\bSigma - \bD\bdelta_A\bdelta_A^T\bD
                \end{array}\right], \nu\right).
\nonumber
\end{eqnarray}
In this parameterization,
the scale matrix $\bSigma$ is partitioned into $\bD\bR\bD$,
making the skewness parameter invariant to a change of scale.
Setting $\bdelta$ in (\ref{BrancoST}) to $\bD\bdelta_A$
leads to the B-MST distribution (\ref{AzzaST}).
This characterization of the rMST distribution was considered
by \citet{J028} to define a skew $t$-mixture model,
and an algorithm for parameter estimation was formulated
using a Bayesian framework.
\newline

\noindent\textbf{The skew $t$-distribution of \citeauthor{J019} (G-MST)}\newline
In \citet{J019}, another version
of the restricted skew $t$-distribution is defined,
starting from the A-MSN distribution of \citet{J001}.
In this parameterization, the scale matrix $\bSigma$
is not factored into the product $\bD\bR\bD$,
and the parameter $\bdelta_A$ is replaced by
$\bD^{-1}\bSigma\bdelta_G$,
leading to a density in a slightly simpler algebraic form,
given by
\begin{equation}
f(\by) = 2 t_p(\by;\bmu,\bSigma,\nu) T_1\left(\bdelta_G^T(\by-\bmu)
	\sqrt{\frac{\nu+p}{\nu+d(\by)}};0,  
    1-\bdelta_G^T\bSigma\bdelta_G,\nu+p\right),
\label{GuptaST}
\end{equation}  
where, as before, $d(\by) = (\by-\bmu)^T\bSigma^{-1}(\by-\bmu)$.
Note that (\ref{GuptaST}) is identical to (\ref{BrancoST})
if we rewrite $\bdelta$ in (\ref{BrancoST}) as $\bSigma\bdelta_G$.
It follows that the stochastic representation of
the G-MST distribution (\ref{GuptaST}) can be expressed as
\begin{eqnarray}
\bY &=& \bmu + \left(\bY_1 \mid Y_0 > 0\right),  \label{eq2.29}
\end{eqnarray}
where
\begin{eqnarray}
\left[\begin{array}{c} Y_0 \\ \bY_1 \end{array}\right]
        \sim t_{1+p} \left(\left[\begin{array}{c} 0 \\
        \bzero \end{array}\right],
        \left[\begin{array}{cc} 1 & \bdelta_G^{T}\bSigma \\
        \bSigma\bdelta_G &
        \bSigma    \end{array}\right], \nu\right).
\label{eq2.30}
\end{eqnarray}
\newline

\noindent\textbf{The skew normal/independent skew $t$-distribution (SNI-ST)}\newline
The skew $t$ member of the SNI class, denoted by SNI-ST,
is introduced as a scale mixture of SNI-SN distributions
with gamma scale factor \citep{J049}.
Its density is given by
\begin{eqnarray}
f(\by) &=& 2 t_p(\by; \bmu, \bSigma,\nu) \nonumber\\
		&&	T_1\left(\bdelta_S^T
            \bSigma^{-\frac{1}{2}} \left(\by - \bmu\right)
            \sqrt{\frac{\nu+p}{\nu+d(\by)}}; 
            0, 1-\bdelta_S^T\bdelta_S, \nu+p\right),
\label{SNIST}
\end{eqnarray}
where $d(\by) = (\by-\bmu)^T\bSigma^{-1}(\by-\bmu)$,
and $\bSigma^{\frac{1}{2}}$ is the square root matrix of $\bSigma$;
that is, $\bSigma^{\frac{1}{2}} \bSigma^{\frac{1}{2}} = \bSigma$.
The SNI-ST distribution (\ref{SNIST}) can be generated by
taking $\bY = \bmu + (\bY_1\mid Y_0>0)$, where
\begin{eqnarray}
\left[\begin{array}{c} Y_0 \\ \bY_1 \end{array}\right]
        \sim t_{1+p} \left(\left[\begin{array}{c} 0 \\
        \bzero \end{array}\right],
        \left[\begin{array}{cc} 1
        & \bdelta_S^{T}\bSigma^{\frac{1}{2}} \\
        \bSigma^{\frac{1}{2}}\bdelta_S & \bSigma    
        \end{array}\right], \nu\right),
\label{eq2.31}
\end{eqnarray}
and the corresponding convolution-type representation is given by
\begin{equation}
\bY = \bmu + \bSigma^{\frac{1}{2}}\bdelta_S| \tilde{Y}_0|
    + \bSigma^{\frac{1}{2}}(I_p-\bdelta_S\bdelta_S^T)^{\frac{1}{2}} 
    \tilde{\bY}_1,
\label{eq2.32}
\end{equation}
where $\tilde{Y}_0$ and $\tilde{\bY}_1$
are jointly distributed as
\begin{eqnarray}
\left[\begin{array}{c} \tilde{Y}_0\\\tilde{\bY}_1
                \end{array}\right]
                \sim t_{1+p}\left(     
                \left[\begin{array}{c}0\\\bzero\end{array}\right],
                \left[\begin{array}{cc}1&\bzero 
                \\\bzero&\bI_p\end{array}\right], \nu\right)\nonumber
\end{eqnarray}
It can observed that (\ref{SNIST})
is equivalent to (\ref{BrancoST})
by replacing $\bdelta$ in (\ref{BrancoST})
with $\bSigma^{\frac{1}{2}}\bdelta_S$.
\citet{J047} and \citet{J066} studied, respectively,
finite mixtures of univariate and multivariate SNI-ST distributions,
and derived an ECME algorithm for computing the ML estimates
of the model parameters.
\newline

\noindent\textbf{The skew $t$-distribution of \citeauthor{J004} (P-MST)}\newline
In \citet{J004}, a restricted variant of \citet{J002}'s
skew $t$-distribution was introduced,
and its density is given by
\begin{eqnarray}
f(\by) &=& 2 t_p(\by; \bmu, \bSigma, \nu) \nonumber\\
	&&	 T_1\left(\bdelta^T \bSigma^{-1}
    \left(\by - \bmu\right)\sqrt{\frac{\nu+p}{\nu+d(\by)}}
    ; 0, 1-\bdelta^T\bSigma^{-1}\bdelta,\nu+p\right),
\label{SamST}
\end{eqnarray}
where $d(\by) = (\by-\bmu)^T\bSigma^{-1}(\by-\bmu)$.
We shall refer to the density (\ref{SamST})
as the rMST distribution.
This distribution has straightforward conditioning
and convolution-type stochastic representations, given by
\begin{equation}
\bY = \bmu + \left(\bY_1\mid Y_0 > 0\right),
\nonumber
\end{equation}
and
\begin{equation}
\bY  = \bmu + \bdelta|\tilde{Y}_0| + \tilde{\bY}_1,
\nonumber
\end{equation}
respectively, where
\begin{eqnarray}
\left[\begin{array}{c} Y_0 \\ \bY_1 \end{array}\right]
        \sim t_{1+p} \left(\left[\begin{array}{c} 0 \\
        \bzero \end{array}\right],
        \left[\begin{array}{cc} 1 & \bdelta^{T} \\
        \bdelta & \bSigma    \end{array}\right], \nu\right),
\label{eq2.34}
\end{eqnarray}
and 
\begin{eqnarray}
\left[\begin{array}{c} \tilde{Y}_0\\\tilde{\bY}_1
                \end{array}\right]
                \sim t_{1+p}\left(     
                \left[\begin{array}{c}0\\\bzero\end{array}\right],
                \left[\begin{array}{cc}1&\bzero 
                \\\bzero&\tilde{\bSigma}\end{array}\right], \nu\right),
\nonumber
\end{eqnarray}
and where $\tilde{\bSigma} = \bSigma - \bdelta \bdelta^T$.
It can be observed that the restricted MST (\ref{SamST})
is identical to (\ref{eq2.23}).
Mixtures of rMST distributions was first studied by \citet{J004},
and a closed-form implementation of the EM algorithm was outlined. 
\citet{J077} subsequently provided an alternative exact implementation.

A summary of the correspondence between the parameters
used in various versions of the restricted MST distribution
is given in Table \ref{tab:rMSTc}.
Their densities and stochastic representations are listed
in Tables \ref{tab:rMSTa} and \ref{tab:rMSTb}.
\newline

\begin{table}
    \centering
    \begin{tabular}{ccc}
    \hline
    rMST    &    $\bdelta$    			&    $\bSigma$    \\
    \hline
    B-MST    &    $\bdelta$    			&    $\bSigma$    \\
    A-MST    &    $\bD\bdelta_A$    	&    $\bSigma$    \\
    G-MST    &    $\bSigma\bdelta_G$    &    $\bSigma$    \\
    P-MST    &    $\bdelta$    			&    $\bSigma$    \\
    SI-ST    &    $\bSigma^{\frac{1}{2}}\bdelta_S$    &    $\bSigma$    \\
    \hline
    \end{tabular}
    \caption{Correspondence between the parametrization of
    the restricted forms of MST distributions.}
    \label{tab:rMSTc}
\end{table}

\subsubsection{Unrestricted multivariate skew $t$-distributions}
\label{sec:uMST}

In the unrestricted case, the latent variable $\bY_0$
is a $p$-dimensional random vector following a $t$-distribution.
Under this setting, $\bY$ is given in terms of
the conditional distribution of $\bY_1$ given $\bY_0$ is positive.
The condition $\bY_0 > \bzero $ implies that
each element of $\bY_0$ is greater than zero.
Similar to (\ref{eq2.23}), the unrestricted MST distribution
takes the form
\begin{equation}
\bY = \bmu + \left(\bY_1 \mid \bY_0 > \bzero \right),
\label{eq2.35}
\end{equation}
where
\begin{eqnarray}
\left[\begin{array}{c} \bY_0 \\ \bY_1 \end{array}\right]
        \sim t_{2p} \left(\left[\begin{array}{c} \bzero \\
        \bzero \end{array}\right],
        \left[\begin{array}{cc} \bI_p & \bDelta \\
        \bDelta & \bSigma
        \end{array}\right], \nu \right).
\label{Eq2.36}
\end{eqnarray}

The analogous convolution-type representation is given by
\begin{equation}
\bY = \bmu +  \bDelta |\tilde{\bY}_0| + \tilde{\bY_1},
\label{eq2.37}
\end{equation}
where the two random vectors $\tilde{\bY}_0$
and $\tilde{\bY}_1$ are jointly distributed as
\begin{eqnarray}
\left[\begin{array}{c} \tilde{Y}_0\\\tilde{\bY}_1
                \end{array}\right]
                \sim t_{2p}\left(     
                \left[\begin{array}{c} \bzero\\\bzero\end{array}\right],
                \left[\begin{array}{cc}\bI_p&\bzero 
                \\\bzero&\tilde{\bSigma}\end{array}\right], \nu\right),
\nonumber
\end{eqnarray}
and where $\tilde{\bSigma} = \bSigma - \bDelta^2$. \newline

This form of the MST distribution is studied in detail in \citet{J002},
and its density is given by
\begin{equation}
f(\by) = 2^p \; t_p(\by; \bmu, \bSigma, \nu) \;
    T_p\left(\bDelta\bSigma^{-1}(\by-\bmu)
    \sqrt{\frac{\nu+p}{\nu+d(\by)}}; \bzero, \bLambda, \nu+p\right),
\label{SahuST}
\end{equation}
where $\bDelta = \mbox{diag} \left(\bdelta\right)$,
$\bLambda = \bI_p - \bDelta \bSigma^{-1} \bDelta$, and 
$d(\by) = (\by-\bmu)^T\bSigma^{-1} (\by-\bmu)$.
ML estimation for the unrestricted characterization
of the MST distribution is a difficult computational problem.
\citet{J027} used a Monte Carlo (MC) E-step
when implementing the EM algorithm.
Later, \citet{J068, J067}, and \citet{J103} proposed
an improved implementation using a truncated moments approach.

It is important to point out that,
although the rMST distribution (\ref{SamST})
was originally obtained as a restricted variant
of the uMST distribution (\ref{SahuST}),
and both can be constructed by the conditioning
and convolution approach,
where (\ref{SahuST}) uses a $p$-dimensional latent variable
instead of a scalar latent variable used in (\ref{SamST}),
the density (\ref{SahuST}) does not incorporate (\ref{SamST}).
The two densities are equivalent only in the univariate case.
\newline

\subsubsection{Extended multivariate skew $t$-distributions}
\label{sec:eMST}

There are parallel versions of the ESN and the SUN distributions
for the skew $t$-distribution,
known as the extended skew $t$ (EST) distribution \citep{J074}
and the unified skew $t$ (SUT) distribution \citep{J017}, respectively.
Their links are analogous to those for the skew normal distributions
in Section \ref{sec:eMSN}.

\subsubsection{Generalized multivariate skew $t$-distributions}
\label{sec:gMST}

Similar to the generalized forms of the MSN distribution, 
analogous extensions
to the skew $t$ case can be constructed.
They include the FUST distribution and other subclasses of it,
as well as the 
generalized form of the $t$-distribution put forward by \citet{J071}, 
known as the selection $t$-distribution.

\section{Mixtures of multivariate skew normal and skew \\ $t$-distributions}
\label{sec:FM}

In a mixture model context, the underlying population 
can be conceptualized as being
composed of a finite number of subpopulations.
Let $\bY = {\bY_1, \ldots, \bY_n}$
denote a random sample of $n$ observations.
Then the probability density function (pdf)
of the $g$ component finite mixture model takes the form
\begin{equation}
f(\by; \bPsi) = \sum_{h=1}^g \pi_h f(\by; \btheta_h),
\label{eq3.1}
\end{equation}
where $f(\by; \btheta_h)$ is the density of the $h$th population,
and $\pi_h$ its corresponding weight.
The mixing proportions $\pi_h$ satisfies
$\pi_h \ge 0$ $(h = 1, \ldots, h)$,
and $\sum_{h=1}^g \pi_h = 1$.
The vector $\btheta_h$ consists of the unknown parameters
in the postulated form of the $h$th component of the mixture model,
and \linebreak
$\bPsi = \left(\pi_1, \ldots, \pi_{g-1}, \btheta_q, \ldots, \btheta_g\right)^T$
denotes the vector containing all unknown parameters.
 
Computation of the ML estimates of the model parameters
is typically achieved through the EM algorithm.
Under the EM framework, the observed data vector
is regarded as incomplete, and latent component labels
(and possibly other latent variables as needed) are introduced.
The unobservable component labels $z_{hj}$
are defined as binary indicator variables,
where $z_{hj}$ takes the value of one when observation
$\by_j$ belongs to the $h$th component, and is zero otherwise.
The E-step computes the so-called $Q$-function,
which is the conditional expectation of the log likelihood function
given the observed data, using the current fit for $\bPsi$.
In the M-step, parameters are updated
by maximizing the $Q$-function obtained from the E-step.
The algorithm then proceeds by alternating the E- and M-steps
until its likelihood increases by an arbitrary small amount
in the case of convergence of the sequence of likelihood values.

\subsection{Finite mixtures of multivariate skew normal distributions}
\label{sec:FMMSN}

\begin{table}
    \centering
    \begin{tabular}{ccc}
        \hline
        Model    &    Algorithm    &    References    \\        
        \hline
        \multicolumn{1}{c}{\textbf{rMSN}}    & \multicolumn{1}{c}{}     &    \\    
        \hline
        FM-rMSN    &    traditional EM    &    \citet{J004}    \\    
        FM-SNI-SN    &    traditional EM    & \citet{J066}    \\
        FM-A-MSN    &    Bayesian EM    & \citet{J028}    \\
        \hline
        \multicolumn{1}{c}{\textbf{uMSN}}    & \multicolumn{1}{c}{}     &    \\    
        \hline
        FM-uMSN    &    traditional EM    &    \citet{J048}    \\
        \hline        
    \end{tabular}
    \caption{EM algorithms for fitting restricted
    and unrestricted forms of multivariate skew normal mixture models.}
    \label{tab:FMMSN}
\end{table}

With reference to (\ref{SamSN}),
the density of a $g$-component finite mixture of
restricted multivariate skew normal (FM-rMSN) distributions
is given by
\begin{equation}
f(\by; \bPsi) = \sum_{h=1}^g \pi_h f(\by; \bmu_h, \bSigma_h, \bdelta_h),
\label{eq:FMrMSN}
\end{equation}
where $f(\by; \bmu_h, \bSigma_h, \bdelta_h)$ refers to
the rMSN density (\ref{SamSN}).
At the $(k+1)$th iteration, the E-step requires the computation
of the conditional expectations
\begin{eqnarray}
e_{1,hj}^{(k)} &=& E_{\bPsi^{(k)}}
    \left\{U_j \mid \by_j, z_{hj}=1\right\},
    \label{eq:3.2a} \\
e_{2,hj}^{(k)} &=& E_{\bPsi^{(k)}}
    \left\{U_j^2 \mid \by_j, z_{hj}=1\right\},
    \label{eq:3.2b}
\end{eqnarray}
where $U_j \mid z_{hj}=1 \sim HN(0, 1)$.
Simple closed-form expressions for the E- and M-steps
of the EM algorithm for fitting mixtures of
restricted forms of MSN distributions can be obtained.
\citet{J004, J066}, and \citet{J028} studied, respectively,
finite mixtures of the rMSN, SNI-SN, and A-MSN distributions,
the latter from a Bayesian perspective (see Table \ref{tab:FMMSN}).
The closed-form EM implementations for FM-rMSN and FM-SNI-SN
are available publicly from the \texttt{R} packages
\texttt{EMMIX-skew} \citep{S002} and \texttt{mixsmsn} \citep{S001}.
On closer examination of the EM algorithm
provided by \citet{J004} and \citet{J066},
it is not difficult to show that
their expressions for the E- and M-steps are identical,  
after an appropriate change in the parameterization
as described in Section \ref{sec:rMSN}

For the unrestricted case, \citet{J048} provided an implementation
of the EM algorithm for fitting the FM-uMSN model.
The conditional expectations required at the E-step are equivalent
to (\ref{eq:3.2a}) and (\ref{eq:3.2b}),
except the latent variable $U_j$
is replaced by a multivariate equivalent.
Closed-form expressions were also achieved for the FM-uMSN model.
This, however, inevitably results in higher computational cost.
Whereas (\ref{eq:3.2a}) and (\ref{eq:3.2b}) can be written
in terms of the (univariate) normal distribution function
for the restricted case, the unrestricted case requires
the computation of the multivariate equivalent.

\subsection{Finite mixtures of multivariate skew $t$-distributions}
\label{sec:FMMST}

\begin{table}
    \centering
    \begin{tabular}{ccc}
        \hline
        Model    &    Algorithm    &    References    \\        
        \hline
        \multicolumn{1}{c}{\textbf{rMST}}    & \multicolumn{1}{c}{}     &    \\    
        \hline
        FM-rMST    &    EM with OSL    &    \citet{J004}    \\    
        FM-rMST    &    traditional EM    &    \citet{J077} \\
        FM-SNI-ST    &    ECME    & \citet{J066}    \\
        FM-A-MST    &    Bayesian EM    & \citet{J028}    \\
        \hline
        \multicolumn{1}{c}{\textbf{uMST}}    & \multicolumn{1}{c}{}     &    \\    
        \hline
        FM-uMST    &    MCEM    &    \citet{J048}    \\
        FM-uMST    &    EM with OSL & \citet{J068}    \\
        FM-uMST    &    ECME    &    \citet{J103}    \\
        \hline        
    \end{tabular}
    \caption{EM algorithms for fitting restricted
    and unrestricted forms of multivariate skew $t$-mixture models.}
    \label{tab:FMMST}
\end{table}

The density of a finite mixture of restricted multivariate
skew $t$ (FM-rMST) distributions is given by
\begin{equation}
f(\by; \bPsi) = \sum_{h=1}^g \pi_h f(\by; \bmu_h, \bSigma_h, \bdelta_h, \nu_h),
\label{eq:FMrMST}
\end{equation}
where $f(\by; \bmu_h, \bSigma_h, \bdelta_h, \nu_h)$
refers to the rMST density (\ref{SamST}).
The necessary conditional expectations
required on the E-step at the $(k+1)$th iteration are
\begin{eqnarray}
e_{1,hj}^{(k)} &=& E_{\bPsi^{(k)}}
    \left\{\log(W_j) \mid \by_j, z_{hj}=1\right\},
    \label{eq:3.3a} \\
e_{2,hj}^{(k)} &=& E_{\bPsi^{(k)}}
    \left\{W_j \mid \by_j, z_{hj}=1 \right\},
    \label{eq:3.3b}    \\
e_{3,hj}^{(k)} &=& E_{\bPsi^{(k)}}
    \left\{W_j U_j \mid \by_j, z_{hj}=1\right\},
    \label{eq:3.3c} \\
e_{4,hj}^{(k)} &=& E_{\bPsi^{(k)}}
    \left\{W_j U_j^2 \mid \by_j, z_{hj}=1\right\},
    \label{eq:3.3d}
\end{eqnarray}
where $U_j \mid w_j, z_{hj}=1 \sim HN(0, 1)$
and $W_j \mid z_{hj}=1 \sim \mbox{gamma}(\nu_h/2, \nu_h/2)$.
Simple closed-form expressions for the E- and M-steps
of the EM algorithm for fitting mixtures of
restricted forms of MST distributions can be obtained.
\citet{J004} (c.f. \citet{J063}), \citet{J028, J066},
and \citet{J077} studied, respectively,
finite mixtures of the rMST, A-MST, SNI-ST, and rMST distributions  
(see Table \ref{tab:FMMST}).

In \citet{J103}, it is pointed out that the EM algorithms
for fitting the FM-rMSN distribution
(in particular, the expressions for (\ref{eq:3.3b})-(\ref{eq:3.3d}))
obtained by \citet{J004} and \citet{J077} are equivalent.
More specifically, the former uses expressions
for the  moments of a (univariate) truncated $t$-distribution
to solve (\ref{eq:3.3c}) and (\ref{eq:3.3d}),
and the latter expresses them in terms of hypergeometric functions.
As in the case of the FM-rMSN and FM-SNI-SN distributions,
the expressions (\ref{eq:3.3b})-(\ref{eq:3.3d}) for the FM-SNI-ST model
are identical to that for the FM-rMST model.
The only difference between the two algorithm lies in
the estimation of the degrees of freedom,
where \citet{J004} and \citet{J063}
use a one-step-late (OSL) approach
to compute the conditional expectation (\ref{eq:3.3a}),
while \citet{J066} employ an ECME algorithm.  
However, it should be noted that the ECME algorithm
presented in \citet{J066} assumes the degrees of freedom
to be the same across all components,
whereas such a restriction was not imposed
when applying the algorithm provided by \citet{J004}.
Again, the implementations of the EM algorithm
for fitting the FM-rMST and FM-SNI-ST models are available from
the \texttt{R} packages \texttt{EMMIX-skew} and \texttt{mixsmsn}, respectively.

In the case of the FM-uMST model, \citet{J027} and \citet{J068}
have put forward two versions of an EM algorithm for fitting
mixtures of unrestricted MST distributions.
The former implemented a Monte Carlo (MC) E-step
for calculating the conditional expectations similar
to (\ref{eq:3.3a})-(\ref{eq:3.3d}), but for the unrestricted case.
The latter employed the OSL approach to calculate (\ref{eq:3.3a}),
and expressed (\ref{eq:3.3c}) and (\ref{eq:3.3d})
in terms of moments of the multivariate truncated $t$-distribution.
\citet{J103} have demonstrated that the second approach
leads to significant reduction in computation time
and improvement in accuracy.
They have also sketched an exact implementation
of the EM algorithm for the FM-uMST model,
which results in an ECME implementation
similar to the algorithm provided by \citet{J066}
for the restricted model.
However, even with the closed-form implementation, 
computation of ML estimates of the parameters for the FM-uMST model
can be slow when the dimension of the data $p$ is large,
due to the computationally intensive procedure 
involved in evaluating the moments of a multivariate truncated $t$-variable.
In view of this, \citet{J116} recently proposed a (restricted)
multivariate skew $t$-normal distribution,
where the (univariate) $t$-distribution function in (\ref{AzzaST})
is replaced by a (univariate) normal distribution function.
With this formulation, the computation time is reduced considerably,
where the most computationally intensive part of the E-step
involves only evaluations of the first two moments
of a (univariate) truncated normal variable.

\section{Clustering DLBCL samples}
\label{sec:DLBCL}

To demonstrate the performance of the multivariate skew mixture models
discussed in Section \ref{sec:FM},
we consider the clustering of a trivariate
Diffuse Large B-cell Lymphoma (DLBCL) dataset
provided by the British Columbia Cancer Agency.
The data contain over 3000 cells derived from
the lymph nodes of patients diagnosed with DLBCL.
Each sample was stained with three markers, namely,
CD3, CD5, and CD19. The task is to cluster
the cells into three groups.
Hence, we fit a three-component FM-uMST model to the data.
For comparison, we include the results of
the FM-rMST model and 
two non-elliptically contoured mixture models, namely, 
finite mixture of multivariate normal-inverse-Gaussian
(FM-MNIG) distributions
and finite mixture of multivariate
shifted asymmetric Laplace (FM-MSAL) distributions.

The MNIG distribution is a flexible parametric family
with four parameters \citep{J078}.
Like the skew $t$-distribution,
the MNIG distribution can accommodate skewness
and heavy tails in the data.
Computation of the ML estimates
of the parameters of the model is carried out by the EM algorithm,
with closed-form E- and M-steps involving modified Bessel functions.   
The MSAL distribution is another alternative
to the skew normal and skew $t$-distribution.
As a three-parameter distribution,
the MSAL distribution has parameters that controls its location,
scale, and skewness. The EM algorithm for fitting mixtures
of MSAL distributions is computationally straightforward
compared to that for the FM-MNIG model
and skew mixture distributions \citep{J104}.

\begin{figure}
    \centering
        \includegraphics[width=0.80\textwidth]{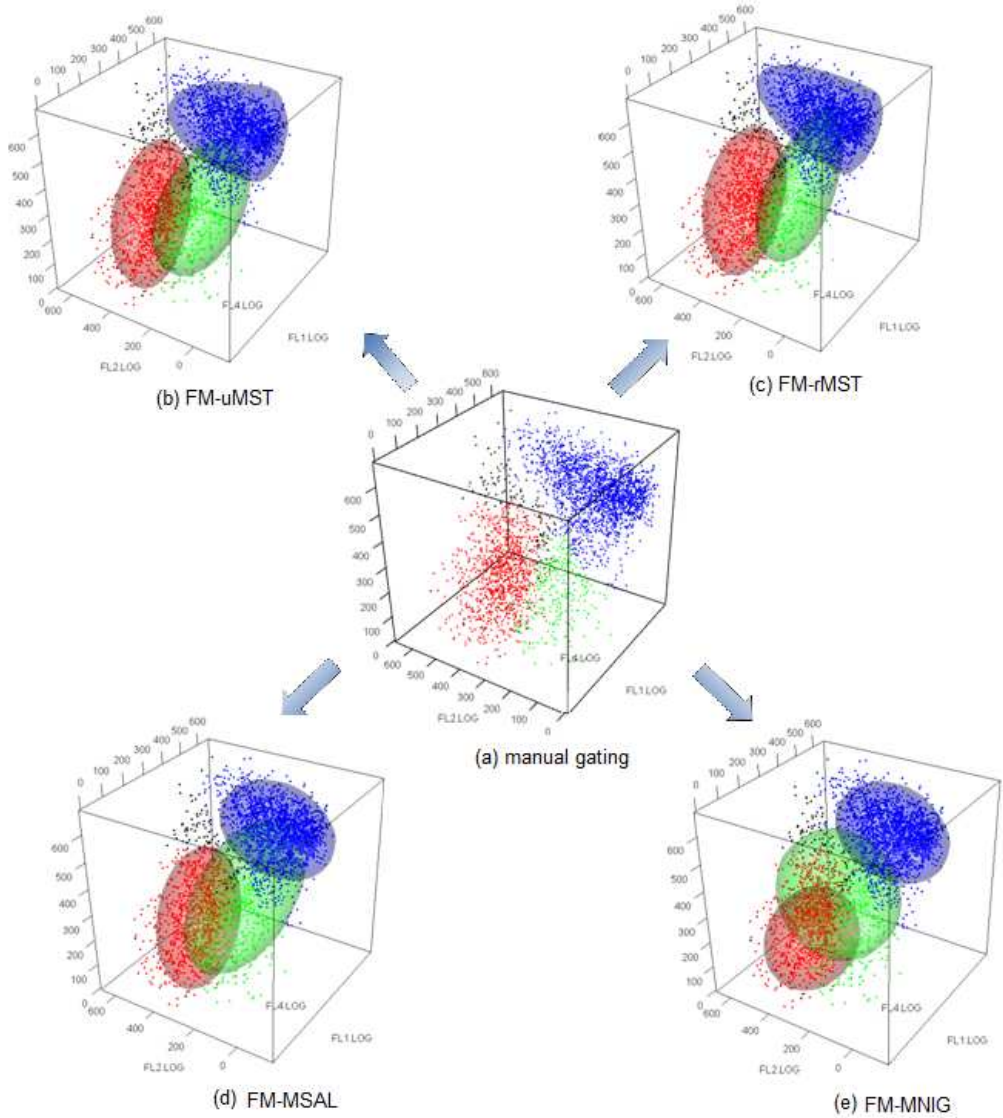}
    \caption{DLBCL dataset: Automated gating results of DLBCL sample
    using four different finite mixture models. The population of
    3290 cells were stained with three fluorescence reagents -
    CD3 (FL1.LOG), CD5 (FL2.LOG), CD19 (FL4.LOG).
    (a) manual expert clustering of the DLBCL into three groups;
    (b) the fitted component contours of the three-component FM-uMST model;
    (c) the contours of the component densities of the fitted restricted
    (FM-rMST) model using \texttt{EMMIX-skew};
    (d) the fitted component contours of the FM-MSAL model;
    (e) the contour plot of the fitted FM-MNIG model.}
    \label{fig1}
\end{figure}

A scatterplot of the data is shown in Figure~\ref{fig1}(a),
where the dots are coloured according to the clustering provided
by human experts, which is taken as the `true' cluster labels.
Figure~\ref{fig1}(b)-(e) shows the density contours
of the components of the fitted FM-uMST, FM-rMST, FM-MNIG,
and FM-MSAL models respectively, which are displayed
with matching colours to Figure~\ref{fig1}(a).
To assess the performance of these algorithms,
we calculated the rate of misclassification
against the `true' results, given by choosing
among the possible permutations of the cluster labels
the one that gives the lowest value.
A lower misclassification or error rate indicates
a closer match between the true labels and the cluster labels
given by the candidate algorithm.
Note that dead cells were removed before
evaluating the misclassification rate.
From Table~\ref{tab:DLBCL1},
it can be seen that
the multivariate skew $t$-mixture models
outperform the other methods in this dataset.
This is also evident in Figure~\ref{fig1},
where the component contours of the FM-uMST and FM-rMST models
resemble quite well the shape of the clusters identified by manual gating.
The results from Table \ref{tab:DLBCL1} reveal that
the unrestricted model is more accurate than the restricted variant
for this dataset.
The FM-MSAL model gives an error rate comparable to that of the FM-rMST model.
However, the FM-MNIG model
has a relatively disappointing performance,
having difficulty in separating the middle (green) and lower (red) clusters.
In future work it is our intention to undertake an extensive comparison of 
the restricted and unrestricted skew $t$-mixture models with mixtures of other  
skew distributions including mixtures of MNIG and MSAL distributions.

\begin{table}
    \centering
        \begin{tabular}{|c|c|c|c|c|}
            \hline
            Model    &    FM-uMST    &    FM-rMST    &    FM-MSAL    &    FM-MNIG \\    
            \hline
            Misclassification rate    & 0.0405 &    0.0638    &    0.0685 &    0.1838    \\
            \hline
        \end{tabular}
    \caption{Misclassification rates for various multivariate
    mixture models on the DLBCL dataset. Cells identified
    as dead cells were not included in the calculation of error rate.}
    \label{tab:DLBCL1}
\end{table}

\section{Concluding Remarks}
\label{sec:concl}

We have presented a schematic way to classify multivariate
skew distributions into four types, namely, the `restricted',
`unrestricted', `extended' and `generalized' forms.
Concerning the use of the terminology `restricted' and `unrestricted',
it should be noted that the restricted skew forms
are not nested within the corresponding unrestricted forms,
with these two forms coinciding in the univariate case.
However, these two forms are both special cases of the extended form,
which itself is a special case of the generalized form.

Current work on finite mixtures of skew distributions
has investigated only the restricted and unrestricted forms
of multivariate skew distributions.
Mixtures based on skew distributions of more general forms
would be of interest for further research.



\begin{thebibliography}{44}
\providecommand{\natexlab}[1]{#1}
\providecommand{\url}[1]{{#1}}
\providecommand{\urlprefix}{URL }
\expandafter\ifx\csname urlstyle\endcsname\relax
  \providecommand{\doi}[1]{DOI~\discretionary{}{}{}#1}\else
  \providecommand{\doi}{DOI~\discretionary{}{}{}\begingroup
  \urlstyle{rm}\Url}\fi
\providecommand{\eprint}[2][]{\url{#2}}

\bibitem[{Arellano-Valle and Azzalini(2006)}]{J017}
Arellano-Valle RB, Azzalini A (2006) On the unification of families of
  skew-normal distributions. Scandinavian Journal of Statistics 33:561--574

\bibitem[{Arellano-Valle and Genton(2005)}]{J008}
Arellano-Valle RB, Genton MG (2005) On fundamental skew distributions. Journal
  of Multivariate Analysis 96:93--116

\bibitem[{Arellano-Valle and Genton(2010)}]{J074}
Arellano-Valle RB, Genton MG (2010) Multivariate extended skew-$t$
  distributions and related families. METRON 68:201--234

\bibitem[{Arellano-Valle et~al.(2006)Arellano-Valle, Branco, and Genton}]{J071}
Arellano-Valle RB, Branco MD, Genton MG (2006) A unified view on skewed
  distributions arising from selections. The Canadian Journal of Statistics
  34:581--601

\bibitem[{Arellano-Valle et~al.(2008)Arellano-Valle, Castro, Genton, and
  G\'omez}]{J076}
Arellano-Valle RB, Castro LM, Genton MG, G\'omez HW (2008) {B}ayesian inference
  for shape mixtures of skewed distributions, with application to regression
  analysis. {B}ayesian Analysis 3:513--540

\bibitem[{Arnold et~al.(1993)Arnold, Beaver, and Meeker}]{J022}
Arnold BC, Beaver RJ, Meeker WQ (1993) The nontruncated marginal of a truncated
  bivariate normal distribution. Psychometrika 58:471--478

\bibitem[{Azzalini(1985)}]{J005}
Azzalini A (1985) A class of distributions which includes the normal ones.
  Scandinavian Journal of Statistics 12:171--178

\bibitem[{Azzalini(2005)}]{J021}
Azzalini A (2005) The skew-normal distribution and related multivariate
  families. Scandinavian Journal of Statistics 32:159--188

\bibitem[{Azzalini and Capitanio(1999)}]{J100}
Azzalini A, Capitanio A (1999) Statistical applications of the multivariate
  skew-normal distribution. Journal of the Royal Statistical Society Series B
  61(3):579--602

\bibitem[{Azzalini and Capitanio(2003)}]{J006}
Azzalini A, Capitanio A (2003) Distribution generated by perturbation of
  symmetry with emphasis on a multivariate skew \emph{t} distribution. Journal
  of the Royal Statistical Society Series B 65(2):367--389

\bibitem[{Azzalini and {\uppercase{D}alla Valle}(1996)}]{J001}
Azzalini A, {\uppercase{D}alla Valle} A (1996) The multivariate skew-normal
  distribution. Biometrika 83(4):715--726

\bibitem[{Basso et~al.(2010)Basso, Lachos, Cabral, and Ghosh}]{J047}
Basso RM, Lachos VH, Cabral CRB, Ghosh P (2010) Robust mixture modeling based
  on scale mixtures of skew-normal distributions. Computational Statistics and
  Data Analysis 54:2926--2941

\bibitem[{Branco and Dey(2001)}]{J012}
Branco MD, Dey DK (2001) A general class of multivariate skew-elliptical
  distributions. Journal of Multivariate Analysis 79:99--113

\bibitem[{Cabral et~al.(2012)Cabral, Lachos, and Prates}]{J066}
Cabral CRB, Lachos VH, Prates MO (2012) Multivariate mixture modeling using
  skew-normal independent distributions. Computational Statistics and Data
  Analysis 56:126--142

\bibitem[{Contreras-Reyes and Arellano-Valle(2012)}]{J119}
Contreras-Reyes JE, Arellano-Valle RB (2012) Growth curve based on scale
  mixtures of skew-normal distributions to model the age-length relationship of
  cardinalfish (epigonus crassicaudus). arXiv:12125180 [statAP]

\bibitem[{Franczak et~al.(2012)Franczak, Browne, and McNicholas}]{J104}
Franczak BC, Browne RP, McNicholas PD (2012) Mixtures of shifted asymmetric
  laplace distributions. arXiv:{12071727} [statME]

\bibitem[{Fr\"{u}hwirth-Schnatter and Pyne(2010)}]{J028}
Fr\"{u}hwirth-Schnatter S, Pyne S (2010) {B}ayesian inference for finite
  mixtures of univariate and multivariate skew-normal and skew-$t$
  distributions. Biostatistics 11:317--336

\bibitem[{Genton(2004)}]{B001}
Genton MG (ed)  (2004) Skew-elliptical Distributions and their Applications: a
  Journey beyond Normality. Chapman \& Hall, CRC, Boca raton, Florida

\bibitem[{Genton and Loperfido(2005)}]{J023}
Genton MG, Loperfido N (2005) Generalized skew-elliptical distributions and
  their quadratic forms. Annals of the Institute of Statistical Mathematics
  57:389--401

\bibitem[{Gonz\'alez-Far\'as et~al.(2004)Gonz\'alez-Far\'as,
  Dom\'inguez-Molinz, and Gupta}]{J018}
Gonz\'alez-Far\'as G, Dom\'inguez-Molinz JA, Gupta AK (2004) Additive
  properties of skew normal random vectors. Journal of Statistical Planning and
  Inference 126:521--534

\bibitem[{Gupta(2003)}]{J019}
Gupta AK (2003) Multivariate skew-$t$ distribution. Statistics 37:359--363

\bibitem[{Gupta et~al.(2004)Gupta, Gonz\'alez-Far\'iaz, and
  Dom\'inguez-Molina}]{J014}
Gupta AK, Gonz\'alez-Far\'iaz G, Dom\'inguez-Molina JA (2004) A multivariate
  skew normal distribution. Journal of Multivariate Analysis 89:181--190

\bibitem[{Ho et~al.(2012)Ho, Lin, Chen, and Wang}]{J067}
Ho HJ, Lin TI, Chen HY, Wang WL (2012) Some results on the truncated
  multivariate $t$ distribution. Journal of Statistical Planning and Inference
  142:25--40

\bibitem[{Iversen(2010)}]{T001}
Iversen DH (2010) Closed-skew distributions: Simulation, inversion and
  parameter estimation. Master's thesis, Norwegian University of Science and
  Technology

\bibitem[{Karlis and Santourian(2009)}]{J078}
Karlis D, Santourian A (2009) Model-based clustering with non-elliptically
  contoured distributions. Statistics and Computing 19:73--83

\bibitem[{Lachos et~al.(2010)Lachos, Ghosh, and Arellano-Valle}]{J049}
Lachos VH, Ghosh P, Arellano-Valle RB (2010) Likelihood based inference for
  skew normal independent linear mixed models. Statistica Sinica 20:303--322

\bibitem[{Lee and McLachlan(2011)}]{J068}
Lee SX, McLachlan GJ (2011) On the fitting of mixtures of multivariate skew
  t-distributions via the {EM} algorithm. arXiv:11094706 [statME]

\bibitem[{Lee and McLachlan(2013)}]{J103}
Lee SX, McLachlan GJ (2013) Finite mixtures of multivariate skew
  $t$-distributions: some recent and new results. Statistics and Computing
  \doi{10.1007/s11222-012-9362-4}

\bibitem[{Lin(2009)}]{J048}
Lin TI (2009) Maximum likelihood estimation for multivariate skew normal
  mixture models. Journal of Multivariate Analysis 100:257--265

\bibitem[{Lin(2010)}]{J027}
Lin TI (2010) Robust mixture modeling using multivariate skew $t$ distribution.
  Statistics and Computing 20:343--356

\bibitem[{Lin et~al.(2007{\natexlab{a}})Lin, Lee, and Hsieh}]{J026}
Lin TI, Lee JC, Hsieh WJ (2007{\natexlab{a}}) Robust mixture modeling using the
  skew-$t$ distribution. Statistics and Computing 17:81--92

\bibitem[{Lin et~al.(2007{\natexlab{b}})Lin, Lee, and Yen}]{J046}
Lin TI, Lee JC, Yen SY (2007{\natexlab{b}}) Finite mixture modelling using the
  skew normal distribution. Statistica Sinica 17:909--927

\bibitem[{Lin et~al.(2013)Lin, Ho, and Lee}]{J116}
Lin TI, Ho HJ, Lee CR (2013) Flexible mixture modelling using the multivariate
  skew-$t$-normal distribution. Statistics and Computing
  \doi{10.1007/s11222-013-9386-4}

\bibitem[{Liseo and Loperfido(2003)}]{J015}
Liseo B, Loperfido N (2003) A {B}ayesian interpretation of the multivariate
  skew-normal distribution. Statistics \& Probability Letters 61:395--401

\bibitem[{Ma and Genton(2004)}]{J072}
Ma Y, Genton MG (2004) A flexible class of skew-symmetric distributions.
  Scandinavian Journal of Statistics 31:459--468

\bibitem[{Prates et~al.(2011)Prates, Lachos, and Cabral}]{S001}
Prates M, Lachos V, Cabral C (2011) \texttt{mixsmsn}: Fitting finite mixture of
  scale mixture of skew-normal distributions.
  \urlprefix\url{http://CRAN.R-project.org/package=mixsmsn}, {R} package
  version 1.0-7

\bibitem[{Pyne et~al.(2009)Pyne, Hu, Wang, Rossin, Lin, Maier, Baecher-Allan,
  McLachlan, Tamayo, Hafler, De~Jager, and Mesirow}]{J004}
Pyne S, Hu X, Wang K, Rossin E, Lin TI, Maier LM, Baecher-Allan C, McLachlan
  GJ, Tamayo P, Hafler DA, De~Jager PL, Mesirow JP (2009) Automated
  high-dimensional flow cytometric data analysis. Proceedings of the National
  Academy of Sciences USA 106:8519--8524

\bibitem[{Riggi and Ingrassia(2013)}]{J118}
Riggi S, Ingrassia S (2013) Modeling high energy cosmic rays mass composition
  data via mixtures of multivariate skew-$t$ distributions. arXiv:13011178
  [astro-phHE]

\bibitem[{Sahu et~al.(2003)Sahu, Dey, and Branco}]{J002}
Sahu SK, Dey DK, Branco MD (2003) A new class of multivariate skew
  distributions with applications to {B}ayesian regression models. The Canadian
  Journal of Statistics 31:129--150

\bibitem[{Soltyk and Gupta(2011)}]{J108}
Soltyk S, Gupta R (2011) Application of the multivariate skew normal mixture
  model with the {EM} algorithm to {Value-at-Risk}. {MODSIM 2011 - 19th
  International Congress on Modelling and Simulation, Perth, Australia,
  December 12-16, 2011}

\bibitem[{Vrbik and McNicholas(2012)}]{J077}
Vrbik I, McNicholas PD (2012) Analytic calculations for the {EM} algorithm for
  multivariate skew $t$-mixture models. Statistics and Probability Letters
  82:1169--1174

\bibitem[{Vrbik and McNicholas(2013)}]{J117}
Vrbik I, McNicholas PD (2013) Parsimonious skew mixture models for model-based
  clustering and classification. arXiv:13022373 [statCO]

\bibitem[{Wang et~al.(2009{\natexlab{a}})Wang, McLachlan, Ng, and Peel}]{S002}
Wang K, McLachlan GJ, Ng SK, Peel D (2009{\natexlab{a}}) \texttt{EMMIX-skew}:
  {EM} Algorithm for Mixture of Multivariate Skew Normal/$t$ Distributions.
  \urlprefix\url{http://www.maths.uq.edu.au/~gjm/mix\_soft/EMMIX-skew}, {R}
  package version 1.0-12

\bibitem[{Wang et~al.(2009{\natexlab{b}})Wang, Ng, and McLachlan}]{J063}
Wang K, Ng SK, McLachlan GJ (2009{\natexlab{b}}) Multivariate skew $t$ mixture
  models: applications to fluorescence-activated cell sorting data. In: Shi H,
  Zhang Y, Bottema MJ, Lovell BC, Maeder AJ (eds) DICTA 2009 (Conference of
  Digital Image Computing: Techniques and Applications, Melbourne), IEEE
  Computer Society, Los Alamitos, California, pp 526--531

\end{thebibliography}
\end{document}